\documentclass[12pt,english]{article}
\usepackage[affil-it]{authblk}
\usepackage{fancyhdr}
\usepackage[top=2.0cm, bottom=2.0cm, left=2.0cm, right=2.0cm]{geometry}
\usepackage{latexsym, amssymb, enumerate, amsmath,amsthm}
\usepackage{color}
\usepackage{xcolor}
\usepackage{tikz}
\usepackage{changebar}
\usepackage{babel}
\usetikzlibrary{decorations.pathreplacing}
\usepackage{tikz}
\usetikzlibrary{decorations}
\usetikzlibrary{decorations.pathmorphing}
\usepackage{url}

    \usepackage{graphicx}
= 0.5\thickmuskip \arraycolsep = 0.3\arraycolsep

\definecolor{dgreen}{rgb}{0,.5,0}
\definecolor{grau}{gray}{.5}
\definecolor{schwarz}{gray}{0}

\newcommand{\reff}[1]{(\ref{#1})}
\newcommand{\ol}[1]{\overline{#1}}

\newcommand{\cg}[1]{\mathcal{#1}}
\newcommand{\av}[1]{\left|#1\right|}

\newcommand{\Ll}[1]{\left\|#1\right\|}
\newcommand{\N}[2]{\left\|#1\right\|_{#2}}
\newcommand{\brkts}[1]{\left(#1\right)}
\newcommand{\ebrkts}[1]{\left[#1\right]}

\newcommand{\brcs}[1]{\left\{#1\right\}}

\newcommand{\pd}[2]{\frac{\partial #1}{\partial #2}}

\newcommand{\bsplitl}[2]{
\begin{equation}
\begin{split}
#1
\end{split}
\label{#2}
\end{equation}}

\newtheorem{thm}{Theorem}[section]

\newtheorem{lem}[thm]{Lemma}

\newtheorem{defn}[thm]{Definition}

\newtheorem{rem}[thm]{Remark}
\DeclareGraphicsExtensions{.png,.eps,.ps}
\begin{document}

\title{Rate of Convergence of Phase Field Equations in Strongly Heterogeneous Media towards
their Homogenized Limit}
\date{\today{}}

%[M. Schmuck, G. A.  Pavliotis, and S. Kalliadasis]{
\author{Markus Schmuck %\href{http://http://www.macs.hw.ac.uk/~ms713/}{Markus Schmuck}%
	\thanks{ \texttt{M.Schmuck@hw.ac.uk} (corresponding author)}%\href{mailto:m.schmuck@hw.ac.uk}{m.schmuck@hw.ac.uk} (corresponding author)}
	}
	\affil{\small
	School of Mathematical and Computer Sciences\\ 
	and the Maxwell Institute for Mathematical Sciences\\
	Heriot-Watt University\\
	EH14 4AS, Edinburgh, UK
	}	
\author{Grigorios A. Pavliotis%
	\thanks{\texttt{g.pavliotis@imperial.ac.uk}}}
	\affil{
	Department of Mathematics\\
	Imperial College London\\
	South Kensington Campus\\
	SW7 2AZ London, UK
	}	
\author{Serafim Kalliadasis%}
	\thanks{\texttt{s.kalliadasis@imperial.ac.uk}}}
	\affil{
	Department of Chemical Engineering\\
	Imperial College London\\
	South Kensington Campus\\
	SW7 2AZ London, UK
	}

\pagestyle{myheadings}
\markright{\emph{M. Schmuck et al.}\hfill Effective macroscopic phase field equations\hfill}
%\fancyhead[CE]{M. Schmuck et al.}
%\fancyhead[CO]{Effective Stokes-Cahn-Hilliard equation}

\makeatletter
\def\@maketitle{%
  \newpage
  \null
  \vskip 2em%
  \begin{center}%
  \let \footnote \thanks
    {\Large\bfseries \@title \par}%
    \vskip 1.5em%
    {\normalsize
      \lineskip .5em%
      \begin{tabular}[t]{c}%
        \@author
      \end{tabular}\par}%
    \vskip 1em%
    {\normalsize \@date}%
  \end{center}%
  \par
  \vskip 1.5em}
\makeatother

\maketitle

\begin{abstract}
We study phase field equations based on the diffuse-interface approximation
of general homogeneous free energy densities showing different local minima
of possible equilibrium configurations in perforated/porous domains. The
study of such free energies in homogeneous environments found a broad
interest over the last decades and hence is now widely accepted and applied in both
science and engineering. Here, we focus on strongly heterogeneous materials
with perforations such as porous media. To the best of our knowledge, we present 
a general formal derivation of upscaled phase field equations for
arbitrary free energy densities and give a rigorous justification by error estimates 
for a broad class of polynomial free energies. The error between the effective macroscopic 
solution of the new upscaled formulation and the solution of the microscopic phase field problem 
is of order $\epsilon^{1/2}$ for a material given characteristic heterogeneity $\epsilon$. Our
new, effective, and reliable macroscopic porous media formulation of general
phase field equations opens new
modelling directions and computational perspectives for interfacial transport in strongly heterogeneous environments. %Our results are applied to wetting dynamics in porous media.

\vspace{0.25cm}

\emph{Keywords:} Free energies, Cahn-Hilliard/Ginzburg-Landau equations, multiscale modeling, homogenization, porous media, wetting, phase transformations
\end{abstract}

\section{Introduction}\label{sec:Intr}
We consider the well-accepted diffuse-interface formulation
\cite{Cahn1958,vanDerWaals1892} for studying the evolution of interfaces
between different phases. Its broad applicability together with increasing
computational power enables its use to new and increasingly complex
scientific and engineering settings such as the computation of transport
equations in porous media \cite{Sahimi2011} which represents a numerically
very demanding, high-dimensional multiscale problem \cite{Jenny2003}. The
purpose of the present work is to rigorously and systematically provide an
analytically and computationally reliable effective macroscopic description
of how multiple phases invade strongly heterogeneous media such as porous
materials for instance.

We consider the abstract energy density
\bsplitl{
e(\phi)
    := \frac{1}{\lambda}F(\phi) %\frac{\alpha}{2}\phi^2(1 -\frac{\beta}{2\alpha}\phi^2)
    +\frac{\lambda}{2}\av{\nabla\phi}^2\,,
}{FrEn}
where $\phi:=\frac{c_\beta}{c_\alpha+c_\beta}$ is a reduced order parameter
representing the fraction of a single species $\beta$ in a binary solution containing species
$\alpha$ and $\beta$ with densities $c_\alpha$ and $c_\beta$, respectively. The gradient term
$\lambda^2\av{\nabla\phi}^2$ penalizes the interfacial area between
these phases, and $F$ is defined as the general
(Helmholtz) free energy density
\bsplitl{ F(\phi)
    & := U-TS\,,
}{hFrEn}
where $U$ stands for the internal energy, $T$ the temperature, and $S$ the
entropy. Important examples for applications include the \emph{regular solution theory} (also
known as the Flory-Huggins energy density \cite{Flory1942})
\bsplitl{
F(\phi)
	&
	:= R(\phi)
	-TS_I(\phi)
	\,,
}{RegSol}
where $S_I(\phi) := -k_B\ebrkts{
	\phi{\rm ln}\,\phi-(1-\phi){\rm ln}\,(1-\phi)
}$
is the entropy of mixing for ideal solutions and the regular solution term
$R(\phi):=z\omega\phi(1-\phi)$
accounts for the interaction energy between different species. The variable $z$ is
the coordination number defining the number of bonds of $\beta$ with
neighbouring species. $\omega:=\epsilon_{\alpha\alpha}+\epsilon_{\beta\beta}-2\epsilon_{\alpha\beta}$
is the interaction energy parameter accounting for the minima $\epsilon_{\alpha\alpha}$,
$\epsilon_{\beta\beta}$, and $\epsilon_{\alpha\beta}$ of interaction potentials which define attractive and repulsive forces between the species $\alpha$ and $\beta$.

The regular solution theory is of great importance in many different contexts
such as ionic melts \cite{Hillert1970}, water sorption in porous solids
\cite{Bazant2012,Bazant2012a}, and micellization in binary surfactant
mixtures \cite{Huang1997}. Also wetting phenomena, of great interest in
technological applications, especially motivated by recent developments in
micro-fluidics, are often studied using classical sharp-interface
approximations, e.g.~\cite{Nikos2010,Nikos2011a,Nikos2011b}, but also enjoy a
wide-spread use of phase-field
modeling~\cite{Pomeau2001,Yue2010,Wylock2012,David2013} also including the
presence of an electric field (electrowetting, e.g.~\cite{Eck2009,
Nochetto2013}). Furthermore, transport in an electrochemical system
consisting of an electrolyte and an electrode \cite{Guyer2004}, or immiscible
flows \cite{Liu2003} under a polynomial free energy in the form of the
classical double-well potential, i.e., $W(\phi):=\frac{1}{4}(1-\phi^2)^2$ are
relevant applications. Our formal derivation of upscaled phase equations is valid 
for general free energies but the derivation of error estimates 
is based on free energies to the following

\medskip

{\bf Polynomial Class (PC):} \emph{Admissible free energy densities $F$ in \reff{FrEn} are polynomials
of order $2r-1$, i.e.,
\bsplitl{
f(u)
	= \sum_{i=1}^{2r-1}a_iu^i\,,
	\quad r\in\mathbb{N}\,,
	\quad r\geq 2
	\,,
}{PNfrEn}
with $f(u)=F' (u)$ vanishing at $u=0$, that is,
\bsplitl{
F(u)
	= \sum_{i=2}^{2r}b_iu^i\,,
	\quad ib_i=a_{i-1}\,,
	\quad 2\leq i\leq 2r
	\,,
}{PolyDef}
where the leading coefficient of both $F$ and $f$ is positive, i.e.,
$a_{2r-1}=2rb_{2r}>0$.
}

\medskip

Temam \cite{Temam1997} established well-posedness of the Cahn-Hilliard equation for free
energies of class (PC).
Phase-field energy functionals are also of interest in image processing such as inpainting, see
e.g.~\cite{Bertozzi2007}. We note that for computational stability one often replaces the regular solution energy density \reff{RegSol} by
the polynomial double-well potential $W(\phi)$.

In difference to the approach in \cite{Schmuck2012b}, we provide here an
upscaling strategy that is valid for general homogeneous free energy
densities due to the application of a functional Taylor expansion around the
effective upscaled solution and hence provides a convenient methodology for
the homogenization of nonlinear problems. Moreover, we present here, to the
best of our knowledge for the first time, error estimates between the
solution of the microscopic phase field equations solved in a periodic porous
medium and the solution of the correspondingly homogenized/upscaled equations
by Theorem \ref{thm:ErEs}. In the remaining part of this section, we
introduce the basic setting where we want to study the dynamics of
interfaces.

\medskip

{\bf (a) Homogeneous domains $\Omega$.} In the Ginzburg-Landau/Cahn-Hilliard formulation, the total energy is defined by
$E(\phi):=\int_\Omega e(\phi)\,d{\bf x}$ with density \reff{FrEn}
on a bounded domain $\Omega\subset\mathbb{R}^d$ with smooth boundary $\partial\Omega$ and $1\leq d\leq 3$ denotes the spatial dimension. In general, the local
minima of $F$ correspond to the equilibrium limiting values of
$\phi$ representing different phases separated by a diffuse
interface whose spatial extension is governed by the gradient term.
It is well accepted that thermodynamic equilibrium can be achieved
by minimizing the energy $E$, here supplemented by a possible boundary
contribution $\int_{\partial\Omega}g({\bf x})\,d{\bf x}$ for $g({\bf x})\in H^{3/2}(\partial\Omega)$ where $do$ denotes the surface measure.
A widely used minimization over time forms the $H^{-1}$-gradient flow with respect to
$E(\phi)$, i.e.,
\bsplitl{
\textrm{(Homogeneous case)}\,\,\,
\begin{cases}
\pd{}{t}\phi
    = {\rm div}\brkts{
    \hat{\rm M}\nabla\brkts{
       \frac{1}{\lambda} f(\phi) %-\phi+\phi^3
        -\lambda\Delta\phi
        }
    }
    & \quad\textrm{in }\Omega_T\,,%:=\Omega\times]0,T[\,,
\\
\nabla_n\phi:= {\bf n}\cdot\nabla\phi
    = g({\bf x})
    & \quad\textrm{on }\partial\Omega_T %:=\partial\Omega\times]0,T[
    \,,
\\
\nabla_n\Delta\phi
    = 0
    & \quad\textrm{on }\partial\Omega_T
    \,,
%\\
%\phi({\bf x},0)
%    = \psi({\bf x})
%    & \quad\textrm{on }\Omega\,,
\end{cases}
}{PhMo}
where $\Omega_T:=\Omega\times]0,T[$, $\partial\Omega_T:=\partial\Omega\times]0,T[$, $\phi$ satisfies the initial condition $\phi({\bf x},0)
    = \psi({\bf x})$, and $\hat{\rm M}=\brcs{{\rm m}_{ij}}_{1\leq i,j\leq d}$ denotes a mobility tensor with real and bounded elements ${\rm m}_{ij}>0$. The gradient flow \reff{PhMo} is 
weighted by the mobility tensor $\hat{\rm M}$, and is referred to as the
Cahn-Hilliard equation. This equation is a model prototype for interfacial dynamics [e.g.
\cite{Fife1991}] and phase transformation [e.g. \cite{Cahn1958}] under homogeneous Neumann boundary conditions,
i.e., $g=0$, and free energy densities $F$. The mean free energy density \reff{FrEn}
can be derived by a thermodynamic limit from lattice gas models of filled and empty
sites for instance. We note that the double-well potential $W$ is
related to the Lennard-Jones potential in the sense of the LMP
(Lebowitz, Mazel and Presutti) theory \cite[]{Presutti2009} but cannot be
exactly reduced to the atomistic Lennard-Jones potential. Finally, we note that we 
applied a scaling with respect to $\lambda$ which allows to identify the 
Hele-Shaw/Mullins-Sekerka problem in the limit $\lambda\to 0$ which is rigorously 
derived in \cite{Alikakos1994} and discussed below.

It is well-known, that formally, the integrated energy density \reff{FrEn} dissipates along
solutions of the gradient flow~\reff{PhMo}, that means,
%\bsplitl{
$E(\phi(\cdot,t))
    \leq
    E(\phi(\cdot,0))
    =:E_0
$, see \reff{BaEnEs} and \reff{BaEnEs1}.
This follows immediately after differentiating \reff{FrEn} with respect to time and using
\reff{PhMo} for $g=0$. Moreover, we emphasize the interesting connection of the Cahn-Hilliard/phase field
equation to the complicated free boundary value problem known as the Mullins-Sekerka problem \cite{Mullins1964} or the
two-phase Hele-Shaw problem \cite{HeleShaw1898}. Inspired by the formal derivation by Pego \cite{Pego1989}, it was rigorously
verified later on by \cite{Alikakos1994,Soner1995} that the chemical potential
$\mu(\phi):=-\lambda\Delta\phi+\frac{1}{\lambda}f(\phi)\,,$ satisfies in the limit $\lambda\to 0$ for $t\in [0,T]$ the following
\bsplitl{
\textrm{Hele-Shaw/Mullins-Sekerka problem:}\qquad
\begin{cases}
\quad
\Delta \mu
	= 0
	&\quad\textrm{in }\Omega\setminus\Gamma_t\,,
\\\quad
{\bf n}\cdot\nabla\mu
	= 0
	&\quad\textrm{on }\partial\Omega\,,
\\\quad
\mu
	= \sigma\kappa
	&\quad\textrm{on }\Gamma_t\,,
\\\quad
v = \frac{1}{2}\ebrkts{{\bf n}\cdot\nabla\mu}_{\Gamma_t}
	&\quad\textrm{on }\Gamma_t\,,
\\\quad
\Gamma_0
	= \Gamma_{00}
	&\quad\textrm{if }t=0\,,,
\end{cases}
}{HSP}
where $\sigma=\int_{-1}^{1}\brkts{\frac{1}{2}\int_0^sf(r)\,dr}^{1/2}\,ds$ is the interfacial tension, $\kappa$ the mean curvature,
$v$ the normal velocity of the interface $\Gamma_t$, ${\bf n}$ the unit outward normal to either $\partial\Omega$ or $\Gamma_t$, and
$\ebrkts{{\bf n}\cdot\nabla\mu}_{\Gamma_t}:={\bf n}\cdot\nabla\mu^+-{\bf n}\cdot\nabla\mu^-$ where $\mu^+:=\mu\,\bigr|_{\Omega^+_t}$ and $\mu^-:=\mu\,\bigl|_{\Omega^-_t}$ and $\Omega^+_t$ and $\Omega^-_t$ denote the exterior and interior of
$\Gamma_t$ in $\Omega$. Herewith, we also have $\phi\to \pm 1$ in $\Omega^\pm_t$ for all $t\in [0,T]$ as $\lambda\to 0$.

\medskip

%In \cite{Burger}, a
%relation between mean curvature and the standard Cahn-Hilliard
%equation, i.e., where the free energy $F$ is chosen as the double
%well potential, is exploited to analyze it with the help of the
%Willmore flow.

\begin{figure}%[htbp]
%\begin{center}
\centering
%\epsffile{rspa_fig1.eps}
%\epsffile[72 72 200 216]{rspa_fig1.eps}
\includegraphics[width=12cm]{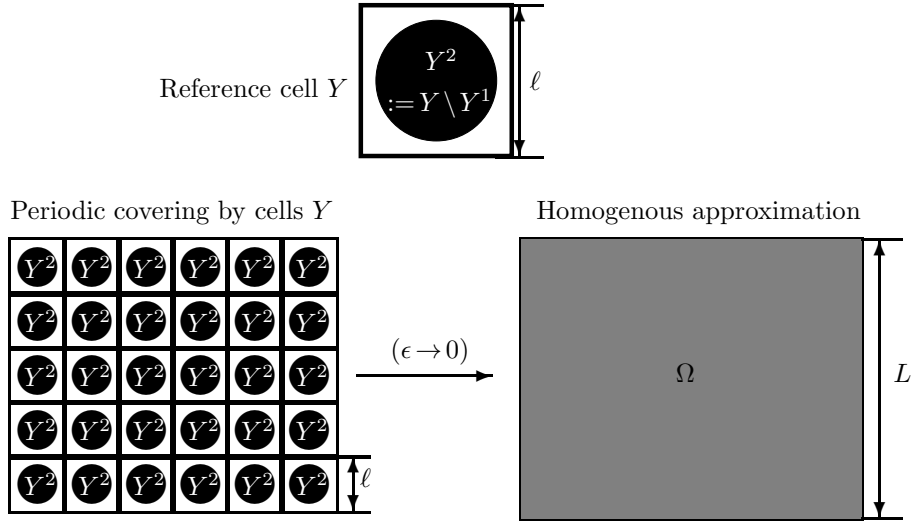}%{rspa_fig1.pdf}
\caption{{\bf Left:} Strongly heterogeneous/perforated material as a periodic covering of reference cells $Y:=[0,\ell]^d$.
{\bf Top, middle:} Definition of the reference cell $Y=Y^1\cup Y^2$ with $\ell=1$.
{\bf Right:} The ``homogenization limit'' $\epsilon:=\frac{\ell}{L}\to 0$ scales the perforated domain such that perforations
become invisible on the macroscale.}
\label{fig:MicMac}
%\end{center}
\end{figure}

{\bf (b) Heterogeneous/perforated domains $\Omega^\epsilon$.} Here we study
the energy density \reff{FrEn} with respect to a perforated domain
$\Omega^\epsilon\subset\mathbb{R}^d$ instead of a homogeneous
$\Omega\subset\mathbb{R}^d$. The dimensionless variable $\epsilon>0$ defines
the heterogeneity $\epsilon =\frac{\ell}{L}$ where $\ell$ represents the
characteristic pore size and $L$ is the characteristic length of the porous
medium, see Figure \ref{fig:MicMac}. Hence, the porous medium is
characterized by a reference cell $Y:=
[0,\ell_1]\times[0,\ell_2]\times\dots\times[0,\ell_d]$ which represents a
single, characteristic pore. For simplicity, we set
$\ell_1=\ell_2=\dots=\ell_d=1$.  A well-accepted approximation is then the
periodic covering of the macroscopic porous medium by such a single reference
cell $\epsilon Y$, see Figure \ref{fig:MicMac}. The pore and the solid phase
of the medium are denoted by $\Omega^\epsilon$ and $B^\epsilon$,
respectively. These sets are defined by, \bsplitl{ \Omega^\epsilon
    & := \bigcup_{{\bf z}\in\mathbb{Z}^d}\epsilon\brkts{Y^1+{\bf z}}\cap\Omega\,,
\qquad
B^\epsilon
    := \bigcup_{{\bf z}\in\mathbb{Z}^d}\epsilon\brkts{Y^2+{\bf z}}\cap\Omega
    =\Omega\setminus\Omega^\epsilon\,,
}{Oe2}
where the subsets $Y^1,\,Y^2\subset Y$ are such that $\Omega^\epsilon$
is a connected set. More precisely, $Y^1$ stands for the pore phase (e.g. liquid or gas phase in wetting problems),
see Figure \ref{fig:MicMac}.

These definitions allow us to reformulate \reff{PhMo}
by the following microscopic porous media problem
\bsplitl{
\textrm{(Micro porous case)}\,\,\,
\begin{cases}
\quad \partial_t\phi_\epsilon
    = {\rm div}\brkts{
    \hat{\rm M}\nabla \brkts{
            -\lambda\Delta \phi_\epsilon
            +\frac{1}{\lambda}f(\phi_\epsilon)
            %-\phi_\epsilon
            %+\phi^3_\epsilon
        }
    }
    & \quad\textrm{in }\Omega^\epsilon_T\,,
\\\quad
\nabla_n\phi_\epsilon:= {\bf n}\cdot\nabla\phi_\epsilon
    = 0
    & \quad\textrm{on }\partial\Omega^\epsilon_T
    \,,
\\\quad
\nabla_n\Delta\phi_\epsilon
    = 0
    & \quad\textrm{on }\partial\Omega^\epsilon_T
    \,,
\\\quad
\phi_\epsilon({\bf x},0)
    = \psi({\bf x})
    & \quad\textrm{on }\Omega^\epsilon\,.
\end{cases}
}{PeMoPr}
In the next section, we motivate our main goal of deriving a homogenized, upscaled problem by passing
to the limit $\epsilon\to 0$ in \reff{PeMoPr}.

\medskip

In Section \ref{sec:2Fo}, we give relevant reformulations of phase field
equations and introduce basic notations and mathematical assumptions. The
main theorems, which state the new effective macroscopic phase field
formulation (Theorem \ref{thm:EfPhFi}) and the associated error (Theorem
\ref{thm:ErEs}) between the solution of the upscaled problem which reliably
accounts for the microstructure by homogenization and the solution of the
microscopic problem fully resolving the pores in Section \ref{sec:MaRe}. The
justification of these results then follow in Sections \ref{sec:FoUp} and
\ref{sec:ErEs}. Conclusions and suggestions for further work are given in 
Section~\ref{sec:Concl}.

\section{Mathematical preliminaries and notation}\label{sec:2Fo}
We present two equivalent formulations of the Cahn-Hilliard
equation. The first helps to achieve solvability for Lipschitz
inhomogeneities and the second, referred to as ``splitting
formulation'', decouples the Cahn-Hilliard equation into two second
order problems for a feasible upscaling by the multiple-scale
method.

\medskip

{\bf (i) Zero mass formulation (for well-posedness)}\label{sec:Ex}
\cite{Novick-Cohen1990} proves well-posedness of a differently scaled Cahn-Hilliard
problem \reff{PhMo} rewritten for $\Omega_T:=\Omega\times]0,T[$ and $\partial\Omega_T:=\partial\Omega\times]0,T[$ in the following zero mass formulation
\bsplitl{
\textrm{\bf (Zero mass)}\quad
\begin{cases}
\quad
\partial_t v
    = {\rm div}\brkts{\hat{\rm M}/\lambda\nabla\brkts{
        b v + h(v) -\lambda\Delta v
    }
    }
    & \quad\textrm{in }\Omega_T\,,%:=\Omega\times]0,T[\,,
\\
\quad
\nabla_n v
    = {\bf n}\cdot\nabla\Delta v
    =0
    & \quad\textrm{on }\partial\Omega_T\,,%:=\partial\Omega\times]0,T[\,,
\\
\quad
v({\bf x},0)
    = v_0({\bf x})
    = \psi({\bf x})-\ol{\phi}\,
    &\quad\textrm{in } \Omega,
\end{cases}
}{ZeMaFo}
where $v({\bf x},t):=\phi({\bf x},t)-\ol{\phi}$, $b:=f'(\phi)$, $h(v):=f(\ol{\phi}+v)-bv$, and by mass conservation of
\reff{PhMo} we define
 %\bsplitl{
$\frac{1}{\av{\Omega}}\int_\Omega \phi\,d{\bf x}
    :=
    \frac{1}{\av{\Omega}}\int_\Omega \psi\,d{\bf x}
    =: \ol{\phi}\,.
$ %}{MaCo}
These definitions imply $bv+h(v)=f(\ol{\phi}+v)$.
We introduce the space
\bsplitl{
H^2_E(\Omega)
    = \brcs{
        \phi\in H^2(\Omega)\,\Bigr|\,\nabla_n\phi=0\textrm{ and }\ol{\phi}=0
    }\,.
}{EnSp}
\cite{Novick-Cohen1990} verifies local existence and uniqueness of solutions
$v\in H^2_E(\Omega)$ of problem \reff{ZeMaFo} for
$f\in C^2_{Lip}(\mathbb{R})$ with $\av{f(s)}\to\infty$ as $s\to\pm\infty$ and
$v({\bf x},0)\in H^2_E(\Omega)$. Moreover, in \cite{Novick-Cohen1990} one also finds necessary
conditions on $h$ leading to global existence.
%Already \cite{Elliott1986} studied
%in the one-dimensional case the existence of blow-up in finite time of \reff{PhMo} with $g=0$, $\lambda>0$, and
%the following polynomial coefficients of $f$, i.e.,
%%\bsplitl{
%$a_3,\,a_2\in\mathbb{R},\textrm{ and }a_1=-1\,.$
% %}{PoCof}
%In fact, they prove that if $a_3<0$, then the solution must blow up in finite time for large
%initial data $\psi$.

%\cite{Novick-Cohen1991,Novick-Cohen1990}
\medskip
{\bf (ii) Splitting (for homogenization)}\label{sec:Sp}
By identifying
%\bsplitl{
$\phi
    = (-\Delta)^{-1}w$
%}{InLap}
in the $H^2_E(\Omega)$-sense, we are able to introduce
the following problem
\bsplitl{
\textrm{\bf (Splitting)}\quad
\begin{cases}
\quad \partial_t(-\Delta)^{-1}w
    -\lambda{\rm div}\brkts{
    \hat{\rm M}\nabla
        w
    }
        = {\rm div}\brkts{
    \frac{\hat{\rm M}}{\lambda}\nabla
        f(\phi) %-\phi+\phi^3
    }
    & \textrm{in }\Omega_T\,,
%\\\quad
%{\bf n}\cdot{\bf J}
%   = F({\bf x})
%   &\textrm{on }\partial\Omega\,,
\\\quad
\nabla_n w
    = -\nabla_n\Delta\phi
    = 0
    &\textrm{on }\partial\Omega_T\,,
\\\quad
-\Delta \phi
    = w
    &\textrm{in }\Omega_T\,,
\\\quad
\nabla_n \phi
    = g({\bf x})%-\frac{\gamma}{C_h}\brkts{
        %a_1\chi_{\partial\Omega^1_w}
        %+a_2\chi_{\partial\Omega^2_w}
    %}
    &\textrm{on }\partial\Omega_T\,,
\\\quad
\phi({\bf x},0)
    = \psi({\bf x})
    & \textrm{in }\Omega\,,
%\\\quad
%{\bf n}\cdot {\bf J}
%   = 0
%   &\textrm{on }\partial\Omega\setminus\brcs{\partial\Omega_l\cup \partial\Omega_w}\,,
\end{cases}
}{DePhMo0}
which is equivalent to \reff{PhMo} in the $H^2_E$-sense and hence, when $g=0$, is well-posed too, \cite{Novick-Cohen1990}.
The advantage of \reff{DePhMo0} is that it allows to base our upscaling approach
on well-known results from elliptic/parabolic homogenization theory \cite[]{Bensoussans1978,Hornung1997,Pavliotis2008,Zhikov1994}.
Finally, we remark that the splitting \reff{DePhMo0} slightly differs from the strategy applied for computational
purposes in \cite{Barrett1999}, for instance.

\medskip

\medskip

For the derivation of a priori estimates (partially based on \cite{Alikakos1994}), the following general characterization of
the homogeneous free energy $f=F'$ proves to be very useful.

\medskip

{\bf Assumption A:}
\emph{
\begin{itemize}
\item[{\rm\bf (A1)}] $F\in C^4(\mathbb{R})$ satisfies $F(\pm 1)=0$ and $F>0$ elsewhere.
\item[{\rm\bf (A2)}] $F'(u)$ satisfies for some finite $\alpha>2$ and positive constants $k_i>0$, $i=0,\dots,3$,
\bsplitl{
k_0\av{u}^{\alpha-2}-k_1
	\leq F'(u)
	\leq k_2\av{u}^{\alpha-2}+k_3
	\,.
}{f2}
\item[{\rm\bf (A3)}] There exist constants $0<a_1\leq 1$, $a_2>0$, $a_3>0$ and $a_4>0$ such that
\bsplitl{
\brkts{
	F(a)-F(b),a-b
	}
	& \geq a_1 \brkts{
		F'(a)(a-b),a-b
	}
	-a_2\av{a-b}^{2+a_3}
	\qquad\forall\av{a}\leq 2a_4\,,
\\
aF''(a)
	& \geq 0
	\qquad\forall\av{a}\geq a_4\,.
}{f3}
\end{itemize}
}

\medskip

It is straightforward to check that the classical double-well potential $F(x)=(x^2-1)^2/4$ satisfies Assumptions A.
In order to obtain more regular solutions, we make the following assumptions on the initial condition \cite{Alikakos1994}.

\medskip

{\bf Assumption B:}
\emph{There exist uniform constants $m_0,\,\sigma_j>0$, $j=1,2,3$ such that
%\begin{align}
%{\rm (B1)}
%	&\qquad%\hspace{4cm}
%	m_0
%		:= \frac{1}{\av{\Omega}}\int_\Omega \phi_0({\bf x})\,d{\bf x}\in ]-1,1[\,,
%	&
%	\nonumber
%\\
%{\rm (B2)}
%	&\qquad
%	F_\lambda(\phi_0)
%		:= \frac{\lambda}{2}\Ll{\nabla \phi_0}^2
%		+\frac{1}{\lambda}\N{F(\phi_0)}{L^1}
%		\leq C\lambda^{-2\sigma_1}
%	&
%	\nonumber
%\\
%{\rm (B3)}
%	&\qquad
%	\N{w_0^\lambda}{H^l}
%	&	
%	\nonumber
%\end{align}
\begin{itemize}
\item[{\rm\bf (B1)}] \qquad $-1 < m_0
		:= \frac{1}{\av{\Omega}}\int_\Omega \psi({\bf x})\,d{\bf x}<1\,,$
\item[{\rm\bf (B2)}] \qquad ${\cal E}_\lambda(\psi)
		:= \frac{\lambda}{2}\Ll{\nabla \psi}^2
		+\frac{1}{\lambda}\N{F(\psi)}{L^1}
		\leq C\lambda^{-2\sigma_1}\,,$
\item[{\rm\bf (B3)}] \qquad $\N{\omega^\lambda}{H^l}
		:=\N{-\lambda\Delta\psi+\frac{1}{\lambda}F(\psi)}{H^l}
		\leq C\lambda^{-\sigma_{2+l}}\,,\quad l=0,1\,.$
\end{itemize}
}

\medskip

Next, we derive some a priori estimates which provide us the necessary bounds for our main result (Theorem \ref{thm:EfPhFi}) quantifying the
error between the solution of the exact microscopic problem and the solution of the upscaled macroscopic equations
(Theorem \ref{thm:ErEs}).

\medskip

\begin{lem} \label{lem:ApEs} \emph{(A priori estimates)} We assume that $f$ and $\psi$ satisfy the \emph{Assumptions A} and \emph{B}.
Moreover, we suppose that the Mullins-Sekerka problem \reff{HSP} has a global in time classical solution. Then, the solution $\phi$ of the
Cahn-Hilliard equation \reff{PhMo} satisfies
\bsplitl{
\N{\phi}{L^\infty(\Omega_T)}
	\leq C
	\,,
}{PhiBnd}
for all $\lambda\in]0,\kappa[$ and a family of smooth initial data $\brcs{\psi^\lambda}_{0<\lambda\leq 1}$
where $\kappa$ and $C$ are constants.

Moreover, for a solution $\phi$ of \reff{PhMo} the following estimate holds
\bsplitl{
\int_0^\infty \Ll{\nabla\Delta\phi}^2\,dt
	\leq C(\lambda)
	\,,
}{phiH3}
for a constant $C>0$. If for a constant $\kappa>0$ the additional inequality
\bsplitl{
\lim_{s\to 0^+}\Ll{\nabla \partial_t\phi(s)}
	\leq C\lambda^{-\kappa}\,,
}{AdCon}
holds, then the solution of \reff{PhMo} also satisfies for a large enough
constant $C>0$
\bsplitl{
\N{\Delta^2\phi}{L^\infty([0,\infty[;L^2(\Omega))}
	\leq C\lambda^{-C}\,.
}{4thOrd}
\end{lem}

\medskip
\begin{rem} (Hele-Shaw) Existence and uniqueness of classical soltuions for the so-called 
single phase Hele-Shaw problem in bounded domains in $\mathbb{R}^d$ has been proved in 
\cite{Escher1997}. Another proof for the existence of a global in time classical solution for the multi-dimensional Hele-Shaw problem can be found in \cite{Meirmanov2002}.
\end{rem}
\medskip

\begin{proof}
\emph{i) Estimate \reff{PhiBnd}:} A proof based on asymptotic analysis and the construction of approximate solutions can be found in \cite{Alikakos1994}.
\\
\emph{ii) Estimate \reff{phiH3}:} This inequality is based on \reff{PhiBnd} and the following basic energy estimate for the
Cahn-Hilliard equation, i.e.,
\bsplitl{
\frac{d E(\phi)}{dt}
	= \brkts{\nabla_\phi^{L^2}E(\phi),\frac{\partial}{\partial t}\phi}
	= \brkts{\mu,\frac{\partial}{\partial t}\phi}
	= -\Ll{\nabla\mu}^2\,,
}{BaEnEs}
which after integration with respect to time leads to
\bsplitl{
\N{E(\phi)}{L^\infty([0,\infty[)}
	+\int_0^\infty\Ll{\nabla\mu}^2\,dt
	\leq E(\psi)\,.
}{BaEnEs1}
We note that $\mu$ is the chemical potential as introduced before equation \reff{HSP}. 
The operator $\nabla_\phi^{L^2}$ denotes the G\^ateaux derivative of $F$ in the 
$L^2$-sense.
%, that means, $\nabla^{L^2}_uf(u):=\lim_{\epsilon\to 0}\frac{f(u+\epsilon v)-f(u)}{\epsilon v}$ for all $v\in L^2$.
We first rewrite $\mu$ as follows
\bsplitl{
\Delta\phi
	= \frac{1}{\lambda^2} f(\phi)
	-\frac{1}{\lambda}\mu
	\,.
}{muRewr}
Applying on both sides the gradient operator leads to
\bsplitl{
\av{\nabla\Delta\phi}
	& =\av{
	\frac{1}{\lambda^2}\nabla f(\phi)
	-\frac{1}{\lambda}\nabla\mu
	}
\\&
	\leq
	\frac{1}{\lambda^2}C\brkts{
		\av{\phi}^{\alpha-2}
		+1
	}\av{\nabla\phi}
	+\frac{1}{\lambda}\av{\nabla\mu}
	\,.
}{MuRewr1}
Using now \reff{PhiBnd} on the first term on the right-hand side, taking the square of the left- and right-hand
side in \reff{MuRewr1}, and a subsequent integration over space together with Assumption (B2) leads to the desired
result.
\\
\emph{iii) Estimate \reff{4thOrd}:} This assertion is an immediate consequence of the triangle inequality,
the $L^\infty$-boundedness \reff{PhiBnd}, and \reff{phiH3}.
\end{proof}

With extension operator $T_\epsilon$, which extends the solutions
$\phi^\epsilon$ and $w^\epsilon$ of \reff{PeMoPr} defined on the microscopic
domain $\Omega^\epsilon$ to the whole domain $\Omega$, the same estimates
(Lemma \ref{lem:ApEs}) hold for $T_\epsilon\phi^\epsilon$ and $T_\epsilon
w^\epsilon$. However, for convenience we denote these extensions by
$\phi^\epsilon$ and $w^\epsilon$ and skip the extension operator $T_\epsilon$
most of the time. The existence of such an operator
$T_\epsilon\,:\,W^{1,p}(\Omega^\epsilon)\to W^{1,p}_{loc}(\Omega)$ for
$\epsilon>0$ was established in \cite{Acerbi1992} and $T_\epsilon$ is characterized by the
following properties:
\bsplitl{
{\bf (T1)}
	& \qquad
	T_\epsilon u =u\quad\textrm{a.e. in }\Omega^\epsilon\,,
\\	
{\bf (T2)}
	& \qquad
	\int_{\Omega(\epsilon k_0)}\av{T_\epsilon u}^p\,d{\bf x}
	\leq k_1\int_{\Omega^\epsilon}\av{u}^p\,d{\bf x}\,,
\\
{\bf (T3)}
	&\qquad
	\int_{\Omega(\epsilon k_0)}\av{D(T_\epsilon u)}^p\,d{\bf x}
	\leq k_2\int_{\Omega^\epsilon}\av{Du}\,d{\bf x}\,,
}{Teps}
for constants $k_0,\,k_1,\,k_2>0$. Hence, $T_\epsilon$ extends solutions defined 
on the pore space $\Omega$ to the whole domain $\Omega$.
\medskip

\begin{rem}\label{rem:Extension} The estimates of Lemma \ref{lem:ApEs} can be derived
in an analogous manner for the perforated problem \reff{PeMoPr} with the help of the
extension operator $T_\epsilon$.
\end{rem}

\section{Main results}\label{sec:MaRe}
Our main result, i.e., the upscaling/homogenization of general phase field equations (including
the Cahn-Hilliard equation), is based on a scale separation property of the chemical potential.

\begin{defn}\label{def:LoEq} \emph{(Scale separation)} We say that
the macroscopic chemical potential is scale
separated if and only if the upscaled chemical potential
\bsplitl{
\mu_0
	:=
	\frac{1}{\lambda}f(\phi)-\lambda\Delta\phi
	\,,
}{ChPo}
satisfies $\frac{\partial \mu_0}{\partial x_l}=0$ for each $1\leq l\leq d$ on the level of the 
reference cell $Y$ but not on the macroscopic domain $\Omega$.
%{\em (Local equilibrium)}
%We say that the phase-field $\phi$ is in local thermodynamic equilibrium,
%% i.e., in each reference cell $Y$,
%if and only if
%\bsplitl{
%\frac{\delta E(\phi)}{\delta \phi}
%    = \mu(\phi)
%    = f(\phi)
%    -\lambda^2\Delta \phi
%    =
%    {\rm const.}\,,
%%    \textrm{ for each $$ reference cell $Y$\,,}
%}{LoEqCo}
%for each ${\bf x}/\epsilon={\bf y}$ element of the same reference cell $Y$. $\mu$ stands for the chemical potential which is only allowed to vary over the different
%reference cells.
\end{defn}

\begin{rem}\label{rem:LoThEq} Definition \ref{def:LoEq} represents a refined
local equilibrium assumption which takes into account the problem specific multiscale
nature, which appears in the homogenization theory as the slow (macroscopic) 
scale $\bf x$ and the fast (microscopic) scale $\bf y$. That means, the variation of the upscaled 
chemical potential with respect to $\bf x$ is too slow on the microscale in the case of scale separation, i.e., $\epsilon\ll 1$.
\end{rem}

The  state of general conditions of equilibrium of heterogeneous substances
seems to go back to the celebrated work of \cite{Gibbs1876}. The assumption
of local thermodynamic equilibrium can be justified on physical and
mathematical grounds by the assumed separation of macroscopic (size of the
porous medium) and microscopic (characteristic pore size) length scales and
the emerging difference in the associated characteristic timescales. This
kind of equilibrium assumptions are widely applied to a veriety of physical
situations such as diffusion \cite[]{Nelson1999}, colloidal
systems~\cite[]{Ben2012}, and macroscale thermodynamics in porous media
\cite[]{Benn1999}, for instance.
The more recent scale separation assumption of Definition \ref{def:LoEq} emerges as a key
requirement for the mathematical well-posedness of arising cell problems
which define effective transport coefficients in homogenized, nonlinear (and
coupled) problems, e.g. ionic transport in porous media based on dilute
solution theory \cite[]{Schmuck2012,Schmuck2013,Schmuck2012a}. 
These considerations allow us to state the following main result of this study.

\begin{thm}\label{thm:EfPhFi}\emph{(Upscaled Cahn-Hilliard equations)}
%Let $\hat{\rm M}=\brcs{{\rm m}\delta_{ij}}_{1\leq i,j\leq d}$ for ${\rm m}>0$ be
%an isotropic mobility tensor. 
We assume that the scale separation 
in the sense of \emph{Definition \ref{def:LoEq}} holds for the macroscopic 
chemical potential $\mu_0$.
Moreover, suppose that $\psi({\bf x})\in H^2_E(\Omega)$. 
Then, the microscopic porous media formulation \reff{PeMoPr} can be effectively
approximated by the following macroscopic problem,
\bsplitl{
\begin{cases}
\theta_1\pd{\phi_0}{t}
    = 
	{\rm div}\brkts{
	\hat{\rm M}_\phi/\lambda\nabla f(\phi_0)
    }
    +\frac{\lambda}{\theta_1}{\rm div}\brkts{
        \hat{\rm M}_w\nabla \brkts{
            {\rm div}\brkts{
                \hat{\rm D}\nabla \phi_0
            }
        }
    }
    &\textrm{in }\Omega_T\,,
\\
\nabla_n \phi_0
    = {\bf n}\cdot\nabla\phi_0
    = 0
    &\textrm{on }\partial\Omega_T\,,
\\
\nabla_n\Delta \phi_0
    = 0
    &\textrm{on }\partial\Omega_T\,,
\\
\phi_0({\bf x},0)
    = \psi({\bf x})
    &\textrm{in }\Omega\,,
\end{cases}
}{pmWrThm}
where $\theta_1:=\frac{\av{Y^1}}{\av{Y}}$ is the porosity and the porous media correction tensors $\hat{\rm D}:=\brcs{{\rm d}_{ik}}_{1\leq i,k\leq d}$,
$\hat{\rm M}_\phi = \brcs{{\rm m}^\phi_{ik}}_{1\leq i,k\leq d}$ and
$\hat{\rm M}_w=\brcs{{\rm m}^w_{ik}({\bf x})}_{1\leq i,k\leq d}$
 are defined by
\bsplitl{
{\rm d}_{ik}
    & := \frac{1}{\av{Y}}\sum^d_{j=1}\int_{Y^1}\brkts{
        \delta_{ik} - \delta_{ij}\pd{\xi^k_\phi}{y_j}
        }
    \,d{\bf y}\,,
\\
{\rm m}^\phi_{ik}
    & :=
    \frac{1}{\av{Y}}\sum_{j=1}^d\int_{Y^1}\brkts{
        {\rm m}_{ik}
        -{\rm m}_{ij}\pd{\xi^k_\phi}{y_j}
    }\,d{\bf y}\,,
\\
{\rm m}^w_{ik}({\bf x})
    & :=
    \frac{1}{\av{Y}}\sum_{j=1}^d\int_{Y^1}\brkts{
        {\rm m}_{ik}
        -{\rm m}_{ij}\pd{\xi^k_w}{y_j}
    }\,d{\bf y}\,.
}{Dik}
The corrector functions $\xi^k_\phi\in H^1_{per}(Y^1)$ and $\xi^k_w\in L^2(\Omega;H^1_{per}(Y^1))$ for $1\leq k\leq d$
solve in the distributional sense the following reference cell problems
\bsplitl{
\xi_w^k:\quad
\begin{cases}
-\sum_{i,j,k=1}^d
    \pd{}{y_i}\brkts{
        {\rm m}_{ik}-{\rm m}_{ij}\pd{\xi^k_w}{y_j}
    }
%\\\qquad\qquad
    =
    -\sum_{k,i,j=1}^d\pd{}{y_i}\brkts{
        {\rm m}_{ik}
        -{\rm m}_{ij}\pd{\xi^k_\phi}{y_j}
    }
    &\textrm{ in }Y^1\,,
\\
\sum_{i,j,k=1}^d{\rm n}_i\Bigl(
        \brkts{
        {\rm m}_{ij}\pd{\xi^k_w}{y_j}
        -{\rm m}_{ik}
        }
	+\brkts{
            {\rm m}_{ik}
            -{\rm m}_{ij}\pd{\xi^k_\phi}{y_j}
        }
    \Bigr)
    = 0
    &\textrm{ on }\partial Y^1_w\cap\partial Y^2_w\,,
\\
\xi^k_w({\bf y})\textrm{ is $Y$-periodic and ${\cg M}_{Y^1}(\xi^k_w)=0$,}
\end{cases}
\\
\xi_\phi^k:\quad
\begin{cases}
-\sum_{i,j=1}^d
    \pd{}{y_i}\brkts{
        \delta_{ik}-\delta_{ij}\pd{\xi^k_\phi}{y_j}
    }
    = 0
    &\textrm{ in }Y^1\,,
\\
\sum_{i,j=1}^d{\rm n}_i
	\brkts{
        \delta_{ij}\pd{\xi^k_\phi}{y_j}
        -\delta_{ik}
        }
    =
    0
    &\textrm{ on }\partial Y^1\,,
\\
\xi^k_\phi({\bf y})\textrm{ is $Y$-periodic and ${\cg M}_{Y^1}(\xi^k_\phi)=0$.}
\end{cases}
}{pmRCTh}
\end{thm}

\begin{rem}\label{rem:Thm}
i) The reference cell problem \reff{pmRCTh}$_1$ for $\xi^k_w$ seems not to allow for analytical 
solutions and can be solved numerically for instance. Moreover, if we assume an isotropic 
mobility, i.e., $\hat{\rm M}:= m\hat{\rm I}$ where $\hat{\rm I}$ is the identity matrix, then 
$\xi^k_w=\xi^k_\phi$ as it solves the same cell problem \reff{pmRCTh}$_2$ as $\xi^k_\phi$.
We note that in this case one also finds results in the literature, e. g. \cite{Auriault1997}, where 
the case of straight or perturbed straight channels is studied. \\
ii) The scale separation property \emph{(Definition \ref{def:LoEq})} of the macroscopic chemical potential $\mu_0$ enables the derivation of the 
cell problem \reff{pmRCTh}$_1$ 
\hfill$\diamond$
\end{rem}

The next result characterizes qualitatively the homogenized Cahn-Hilliard/Ginzburg-Landau phase field equations with the help of error estimates. However, since the microscopic equations 
are defined on the perforated domain $\Omega^\epsilon$ and the upscaled/homogenized 
problem on the whole domain $\Omega$, we will subsequently use the extension 
operator $T_\epsilon$ introduced in \reff{Teps}. For convenience, we will 
understand the variables $\phi^\epsilon$ and $w^\epsilon$ as the extensions 
to $\Omega$ by $T_\epsilon$ in the context of the error estimates, i.e., we do not explicitely write $T_\epsilon\phi^\epsilon$ 
and $T_\epsilon w^\epsilon$, respectively. 

\begin{thm}\label{thm:ErEs} \emph{(Error estimates)} Let $\phi^\epsilon$ be a solution of \reff{PeMoPr} and $w^\epsilon$ the corresponding solution obtained  via substitution as 
in the splitting formulation \reff{DePhMo0}. Let $\hat{\rm M}=m\hat{\rm I}$ be an isotropic mobility with $\hat{\rm I}$ representing the identity matrix.
If the admissible free energy $F$ is polynomial of \emph{class (PC)} satisfying \emph{Assumption A} and $T_\epsilon\xi^k_\phi,\,T_\epsilon\xi^k_w\in W^{1,\infty}_{per}(Y)$ (where we subsequently 
skip the extension operator $T_\epsilon$ for convenience), then the error variables
\bsplitl{
{\rm E}^\phi_\epsilon
	& := \phi^\epsilon-(\phi_0+\epsilon\phi_1)
	\,,
\\
{\rm E}^w_\epsilon
	& := w^\epsilon-(w_0+\epsilon w_1)
\,,
}{ErVar}
where 
$\phi_1:=-\sum_{k=1}^d \xi^k_w({\bf y})\pd{w_0}{x_k}({\bf x},t)$ 
and 
$w_1:=-\sum_{k=1}^d \xi^k_\phi({\bf y})\pd{\phi_0}{x_k}({\bf x},t)$, 
satisfy the following estimates
\bsplitl{
\N{{\rm E}_\epsilon^w(\cdot,T)}{L^2(\Omega)}^2
	+2\brkts{m\lambda^2-5\kappa}
	 \int_0^T\N{\Delta {\rm E}_\epsilon^w(\cdot,t)}{L^2(\Omega)}^2\,dt
%\\&
&	\leq 
		C\brkts{
			\epsilon+\epsilon^{5/2}
		}C(T,\Omega,\kappa)
%		\int_0^T
%			{\rm exp}\bigl(
%				B(T)-B(t)
%			\bigr)
%		\,dt
	\,,
\\
\N{{\rm E}_\epsilon^\phi(\cdot,T)}{L^2(\Omega)}^2
	&\leq  
		C\epsilon \biggl(
			(1+\epsilon^{3/2})C(T,\Omega,\kappa)
%			\int_0^T 
%				{\rm exp}\bigl(B(T)-B(t)\bigr)
%			\,dt
%\\&\quad
			+\epsilon+1%\epsilon(1+\epsilon^{-\frac{1}{2}})^2
		\biggr)
	\,,
}{ThmErEs}
where $C(T,\Omega,\kappa)$ is a constant independent of $\epsilon$.
\end{thm}

We do not expect these error estimates to be optimal.

\section{Proof of Theorem \ref{thm:EfPhFi}}\label{sec:FoUp}
We define the micro-scale $\frac{{\bf x}}{\epsilon}=:{\bf y}\in Y$ such that
the following multiscale properties for spatial differentiation hold, that is, 
$%\bsplitl{
\pd{f_\epsilon({\bf x})}{x_i}
    = \frac{1}{\epsilon}\pd{f}{y_i}({\bf x}, {\bf x}/\epsilon)
    +\pd{f}{x_i}({\bf x}, {\bf x}/\epsilon)\,,
\textrm{ and }%\\
\nabla f_\epsilon ({\bf x})
    = \frac{1}{\epsilon}\nabla_y({\bf x},{\bf x}/\epsilon)
    +\nabla_x({\bf x}, {\bf x}/\epsilon)\,,
$%}{mscLp} 
\,\,where $f_\epsilon({\bf x})=f({\bf x},{\bf  y})$ is an arbitrary function depending on two variables ${\bf x}\in\Omega$,
${\bf y}\in Y$. The Laplace operators $\Delta$ and ${\rm div}\brkts{\hat{\rm M}\nabla}$ then behave as follows,
\bsplitl{
\begin{cases}
{\cg A}_0
   & =
    -\sum_{i,j=1}^d\pd{}{y_i}\brkts{\delta_{ij} \pd{}{y_j}}
    \,,
\\
{\cg A}_1
    &=
    -\sum_{i,j=1}^d\biggl[\pd{}{x_i}\brkts{\delta_{ij}\pd{}{y_j}}
        +\pd{}{y_i}\brkts{\delta_{ij}\pd{}{x_j}}
        \biggr]\,,
\\
{\cg A}_2
   & = - \sum_{i,j=1}^d\pd{}{x_j}\brkts{\delta_{ij}\pd{}{x_j}}\,,
\\
{\cg B}_0
   & =
    - \sum_{i,j=1}^d\pd{}{y_i}\brkts{{\rm m}_{ij}\pd{}{y_j}}
    \,,
\\
{\cg B}_1
   & =
    -\sum_{i,j=1}^d\biggl[\pd{}{x_i}\brkts{{\rm m}_{ij}\pd{}{y_j}}
        +\pd{}{y_i}\brkts{{\rm m}_{ij}\pd{}{x_j}}
        \biggr]\,,    
\\
{\cg B}_2
    & = - \sum_{i,j=1}^d\pd{}{x_j}\brkts{{\rm m}_{ij}\pd{}{x_j}}\,,
\end{cases}
}{B0B1B2}
such that we can define ${\cg A}_\epsilon := \epsilon^{-2}{\cg A}_0
+ \epsilon^{-1}{\cg A}_1 +{\cg A}_2$ and correspondingly ${\cg B}_\epsilon$. Hence, it holds for the
Laplace operator that
$\Delta f_\epsilon({\bf x}) =
{\cg A}_\epsilon f({\bf x}, {\bf y})$. 
To account for the multiscale nature of strongly heterogeneous environments \cite{Schmuck2012,Schmuck2012a,Schmuck2013}, we make the ansatz of 
formal asymptotic expansions
\bsplitl{
w^\epsilon
    & \approx w_0({\bf x},{\bf y},t)
    +\epsilon w_1({\bf x},{\bf y},t)
    +\epsilon^2 w_2({\bf x},{\bf y},t)\,,
\\
\phi^\epsilon
    & \approx \phi_0({\bf x},{\bf y},t)
    +\epsilon \phi_1({\bf x},{\bf y},t)
    +\epsilon^2 \phi_2({\bf x},{\bf y},t)\,,
}{AsExp}
where we neglected higher order terms. 
Before we can insert \reff{AsExp} into the microscopic formulation \reff{DePhMo0}, we need to approximate the derivative of the 
nonlinear homogeneous free energy $f:=F'$ by a Taylor expansion of the form
\bsplitl{
%f(\phi^\epsilon)\approx
	f(\phi^\epsilon)
	\approx
	f(\phi_0)+ f'(\phi_0)(\phi^\epsilon-\phi_0)
	+\frac{1}{2}f''(\phi_0)(\phi^\epsilon-\phi_0)^2
	+ {\cal O}\brkts{(\phi^\epsilon-\phi_0)^3}\,,
}{FTE}
where $\phi_0$ stands for the leading order term in \reff{AsExp}$_2$.\footnote{
Here, we allow for general free energy densities in difference to the subsequent rigorous derivation 
of error estimates which is based on energy densities of the polynomial class (PC).
}
Entering with \reff{AsExp} and \reff{FTE} into \reff{DePhMo0} and using \reff{B0B1B2} provides
the following sequence of problems after equating terms of equal power
in $\epsilon$,
\bsplitl{
\mathcal{O}(\epsilon^{-2}):\quad
\begin{cases}
\mathcal{B}_0\ebrkts{
		\lambda w_0
		+1/\lambda f(\phi_0)
	}
    = 0
    &\textrm{in }Y^1\,,
\\\quad
\textrm{no flux b.c.}\,,
\\\quad
\textrm{$w_0$ is $Y^1$-periodic}\,,
\\
\mathcal{A}_0\phi_0=0
    &\textrm{in }Y^1\,,
\\\quad
\nabla_n \phi_0
    = 0
    &\textrm{on }\partial Y^1_w\cap\partial Y^2_w\,,
\\\quad
\textrm{$\phi_0$ is $Y^1$-periodic}\,,
\end{cases}
}{O-2n}
\bsplitl{
\mathcal{O}(\epsilon^{-1}):\quad
\begin{cases}
{\cg B}_0 \ebrkts{
		\lambda w_1
		+1/\lambda f'(\phi_0)\phi_1
	}
    = -{\cg B}_1\ebrkts{ 
		\lambda w_0
		+1/\lambda f(\phi_0)
	}
 %   -{\cg B}_0\ebrkts{
 %	f'(\phi_0)\phi_1
 %       %f(\phi_0)\frac{\phi_1}{\phi_0}
 %   }
%    -{\cg B}_1f(\phi_0)
    &\textrm{in }Y^1\,,
\\\quad
\textrm{no flux b.c.}\,,
\\\quad
\textrm{$w_1$ is $Y^1$-periodic}\,,
\\
{\cg A}_0 \phi_1
    = -{\cg A}_1 \phi_0
    &\textrm{in }Y^1\,,
\\\quad
\nabla_n \phi_1
    = 0
    &\textrm{on }\partial Y^1_w\cap\partial Y^2_w\,,
\\\quad
\textrm{$\phi_1$ is $Y^1$-periodic}\,,
\end{cases}
}{O-1n}
\bsplitl{
\mathcal{O}(\epsilon^{0}):\quad
\begin{cases}
{\cg B}_0 \ebrkts{
		\lambda w_2
		+\frac{1}{\lambda}\brkts{
			\frac{1}{2}f''(\phi_0)\phi_1^2+f'(\phi_0)\phi_2
		}
	}
	&
\\ \qquad\qquad\qquad\quad
    = -\brkts{
        {\cg B}_2\ebrkts{
		 \lambda w_0
		+1/\lambda f(\phi_0)
	}
        +{\cg B}_1\ebrkts{
		\lambda w_1
		-1/\lambda f'(\phi_0)\phi_1
	}
	}
	&
\\ \qquad\qquad\qquad\quad
%    -{\cg B}_1\ebrkts{
%	f'(\phi_0)\phi_1
%    }
%    -{\cg B}_2 f(\phi_0)
    -\partial_t{\cg A}_2^{-1}w_0
    &\textrm{in }Y^1\,,
\\\quad
\textrm{no flux b.c.}\,,
\\\quad
\textrm{$w_2$ is $Y^1$-periodic}\,,
\\
{\cg A}_0 \phi_2
    = -{\cg A}_2 \phi_0
    -{\cg A}_1 \phi_1
    +w_0
    &\textrm{in }Y^1\,,
\\\quad
\nabla_n \phi_2
    = g_\epsilon
    &\textrm{on }\partial Y^1_w\cap\partial Y^2_w\,,
\\\quad
\textrm{$\phi_2$ is $Y^1$-periodic}\,.
\end{cases}
}{O-0n}

The first problem \reff{O-2n} is of the well-known form from
elliptic homogenization theory \cite{Bensoussans1978,%Cioranescu2000,Marchenko2006,Chechkin2007,
Pavliotis2008} and immediately implies that the
leading order approximation is independent of the microscale ${\bf
y}$. This fact and the linear structure of \reff{O-1n} suggests the
following ansatz for $w_1$ and $\phi_1$, i.e., 
\bsplitl{
w_1({\bf x},{\bf y},t)
    & = -\sum_{k=1}^d \xi^k_w({\bf y})\pd{w_0}{x_k}({\bf x},t)\,,
\\
\phi_1({\bf x},{\bf y},t)
    & %=-\sum_{k=1}^d\xi^k_\phi({\bf y})\frac{\partial v_0}{\partial x_k}({\bf x},t)
    = -\sum_{k=1}^d \xi^k_\phi({\bf y})\pd{\phi_0}{x_k}({\bf x},t)
    =\phi_1({\bf x},{\bf y},t)
    \,.
}{XiwXiphi}

Inserting \reff{XiwXiphi} into \reff{O-1n}$_2$ provides an equation for the correctors $\xi^k_w$ and $\xi^k_v$. The resulting equation for $\xi^k_v$ is again
standard in elliptic homogenization theory and can be immediately written
for $1\leq k\leq d$ as,
\bsplitl{
\xi_\phi:\quad
\begin{cases}
-\sum_{i,j=1}^d
    \pd{}{y_i}\brkts{
        \delta_{ik}-\delta_{ij}\pd{\xi^k_\phi}{y_j}
    }
    =
    &
\\\qquad\quad\,\,
    = -{\rm div}\brkts{
        {\bf e}_k-\nabla_y\xi^k_\phi
    }=0
    &\textrm{ in }Y^1\,,
\\
%\sum_{i,j,k=1}^d{\rm n}_i
%       \delta_{ij}\pd{\xi^k_v}{y_j}
%   =
        {\bf n}\cdot\brkts{
            \nabla\xi^k_\phi
            +{\bf e}_k
        }
    =
    0
    &\textrm{ on }\partial Y^1_w\cap\partial Y^2_w\,,
\\
\xi^k_\phi({\bf y})\textrm{ is $Y$-periodic and ${\cg M}_{Y^1}(\xi^k_\phi)=0$.}
\end{cases}
}{Xiphi}

To study \reff{O-1n}$_1$, we first rewrite ${\cg B}_0\ebrkts{f'(\phi_0)\phi_1}$ and ${\cg B}_1f(\phi_0)$ as follows
\bsplitl{
{\cg B}_0\ebrkts{f'(\phi_0)\phi_1}
	& = -\sum_{k,i,j=1}^d\frac{\partial}{\partial y_i}\brkts{
		{\rm m}_{ij}\frac{\partial\xi^k_\phi}{\partial y_j}\frac{\partial f(\phi_0)}{\partial x_k}
	}\,,
\\
{\cg B}_1 f(\phi_0)
	& = \sum_{i,j=1}^d%\ebrkts{
		\frac{\partial}{\partial y_i}\brkts{
			{\rm m}_{ij}
			\frac{\partial f(\phi_0)}{\partial x_j}
		}\,.
	%}
}{O-1xpl}
Doing the same for $w_1$ and $w_0$ and using \reff{XiwXiphi} leads then to 
\bsplitl{
-\lambda\sum_{k,i,j=1}^d
	\frac{\partial}{\partial y_i}
	&\brkts{
		{\rm m}_{ij}\brkts{
			\frac{\partial x_k}{\partial x_j}
			-\frac{\partial\xi^k_w}{\partial y_j}
		}\frac{\partial w_0}{\partial x_k}
	}
%\\&
	= 1/\lambda\sum_{k,i,j=1}^d
	\frac{\partial}{\partial y_i}\brkts{
		{\rm m}_{ij}\brkts{
			\frac{\partial x_k}{\partial x_j}
			-\frac{\partial\xi^k_\phi}{\partial y_j}
		}\frac{\partial f(\phi_0)}{\partial x_k}
	}\quad\textrm{in $Y^1$}\,.
}{O-1n1}
Next, we assume that the chemical potential $\mu(\phi)=\frac{\delta E(\phi)}{\delta\phi}$ is scale separated, i.e.,
\bsplitl{
\frac{\partial\mu}{\partial x_i}
	=
	\frac{\partial}{\partial x_i}\brkts{
		f(\phi)/\lambda
		-\lambda\Delta\phi
	}
	=
	\frac{\partial}{\partial x_i}\brkts{
		f(\phi)/\lambda
		+\lambda w
	}
	= 0\,.
}{sclsep}
Entering with \reff{sclsep} into \reff{O-1n1} finally gives the reference cell problem for 
$\xi^k_w$, $1\leq k\leq d$ for given $\xi^k_v$
\bsplitl{
\begin{cases}
-\sum_{i,j,k=1}^d
    \pd{}{y_i}\brkts{
        {\rm m}_{ik}-{\rm m}_{ij}\pd{\xi^k_w}{y_j}
    }
\\\qquad\qquad
    =
    -\sum_{k,i,j=1}^d\pd{}{y_i}\brkts{
        {\rm m}_{ik}
        -{\rm m}_{ij}\pd{\xi^k_\phi}{y_j}
    }
    &\textrm{ in }Y^1\,,
\\
\sum_{i,j,k=1}^d{\rm n}_i\Bigl(
        \brkts{
        {\rm m}_{ij}\pd{\xi^k_w}{y_j}
        -{\rm m}_{ik}
        }
%\\\qquad\qquad
        %-\sum_{k,i,j=1}^d%\pd{}{y_i}
	+\brkts{
            {\rm m}_{ik}
            -{\rm m}_{ij}\pd{\xi^k_\phi}{y_j}
        }
    \Bigr)
    = 0
    &\textrm{ on }\partial Y^1_w\cap\partial Y^2_w\,,
\\
\xi^k_w({\bf y})\textrm{ is $Y$-periodic and ${\cg M}_{Y^1}(\xi^k_w)=0$.}
\end{cases}
}{Xiw}

We then come to the last problem \reff{O-0n}. Again, equation
\reff{O-0n}$_2$ is much simpler because it is standard in elliptic
homogenization theory. Well-known existence and uniqueness results
(Fredholm alternative/Lax-Milgram) immediately guarantee solvability
by verifying that the right hand side in \reff{O-0n} is zero as an
integral over $Y^1$. This means, \bsplitl{ -\sum_{i,j=1}^d\int_{Y^1}
    \pd{}{x_i}\brkts{
        \delta_{ij}\brkts{
            \pd{\phi_1}{y_j}
            +\pd{\phi_0}{x_j}
        }
    }\,d{\bf y}
    -\tilde{g}_0
    =\av{Y^1}w_0
    \,,
}{UpPhi0}
where $\tilde{g}_0:=-\frac{\gamma}{C_h}\int_{\partial Y^1}\brkts{a_1\chi_{\partial Y^1_{w_1}}
    +a_1\chi_{\partial Y^1_{w_2}}}\,do({\bf y})$. We obtain the following
effective equation for the phase field,
\bsplitl{
-\sum_{i,k=1}^d\ebrkts{
        \sum_{j=1}^d\int_{Y^1}\brkts{
            \delta_{ik}-\delta_{ij}\pd{\xi^k_\phi}{y_j}
        }\,d{\bf y}
    }\pd{^2\phi_0}{x_i\partial x_k}
    = \av{Y^1}w_0
    +\tilde{g}_0
    \,,
}{UpPhi1}
which can be written more compactly by defining a porous media correction tensor
$\hat{\rm D}:=\brcs{{\rm d}_{ik}}_{1\leq i,k\leq d}$ by
\bsplitl{
\av{Y}{\rm d}_{ik}
    := \sum_{j=1}^d\int_{Y^1}\brkts{
        \delta_{ik}
        -\delta_{ij}\pd{\xi^k_\phi}{y_j}
    }\,d{\bf y}\,.
}{Dphi}
Equations \reff{UpPhi1} and \reff{Dphi} provide the final form of the upscaled
equation for $\phi_0$, i.e.,
\bsplitl{
-\Delta_{\hat{\rm D}} \phi_0
    :=
    -{\rm div}\brkts{
        \hat{\rm D}\nabla \phi_0
    }
    = \theta_1w_0
    +\tilde{g}_0\,.
}{EfPhi0}

The upscaled equation for $w$ is again a result of the Fredholm alternative, i.e., a solvability
criterion on equation \reff{O-0n}$_1$. This means that we require,
\bsplitl{
\int_{Y^1}\Bigl\{
    -\lambda\brkts{
         {\cg B}_2 w_0
        +{\cg B}_1 w_1
    }
%    -{\cg B}_0 \ebrkts{
%        \frac{1}{2}f''(\phi_0)\phi_1^2
%        +f'(\phi_0)\phi_2
%    }
%\\ \qquad\qquad
    -\frac{1}{\lambda}{\cg B}_1\ebrkts{
        f'(\phi_0)\phi_1
    }
    -\frac{1}{\lambda}{\cg B}_2f(\phi_0)
    -\partial_t{\cg A}_2^{-1}w_0
    \Bigr\}\,d{\bf y}
    =0\,.
}{SoCow2}
Let us start with the terms that are easily averaged over the reference
cell $Y$. The first two terms in \reff{SoCow2} can be rewritten by,
\bsplitl{
\int_{Y^1}-\brkts{
        {\cg B}_2 w_0
        +{\cg B}_1 w_1
    }\,d{\bf y}
    =
    \sum_{i,k=1}^d\ebrkts{
        \sum_{j=1}^d\int_{Y^1}\brkts{
            {\rm m}_{ik}-{\rm m}_{ij}\pd{\xi^k_w}{y_j}
        }\,d{\bf y}
    }\pd{^2w_0}{x_i\partial x_k}
\\
    = {\rm div}\brkts{
        \hat{\rm M}_w\nabla w_0
    }
    \,,
}{Bw}
where the effective tensor $\hat{\rm M}_w=\brcs{{\rm m}^w_{ik}}_{1\leq i,k\leq d}$ is defined by
\bsplitl{
{\rm m}^w_{ik}
    & :=
    \frac{1}{\av{Y}}\sum_{j=1}^d\int_{Y^1}\brkts{
        {\rm m}_{ik}
        -{\rm m}_{ij}\pd{\xi^k_w}{y_j}
    }\,d{\bf y}\,.
}{Mw}
The third integrand in \reff{SoCow2} becomes
\bsplitl{
-{\cg B}_1
	&\ebrkts{
		f'(\phi_0)\phi_1
    	}
\\&
	=	
	-\sum_{i,j=1}^d\ebrkts{
		\frac{\partial}{\partial x_i}\brkts{
			{\rm m}_{ij}f'(\phi_0)
			\sum_{k=1}^d\frac{\partial\xi^k_\phi}{\partial y_j}\frac{\partial\phi_0}{\partial x_k}
		}
		+\frac{\partial}{\partial y_i}\brkts{
			{\rm m}_{ij}f'(\phi_0)\sum_{k=1}^d\xi^k_\phi\frac{\partial^2\phi_0}{\partial x_k\partial x_j}
		}
	}\,,
}{Bphi}
where the last term in \reff{Bphi} disappears after integrating by parts. The first term 
on the right-hand side of \reff{Bphi} can be rewritten with the help of the chain rule
\bsplitl{
\frac{\partial^2 f(\phi_0)}{\partial x_k\partial x_j}
	= f''(\phi_0)\frac{\partial\phi_0}{\partial x_k}\frac{\partial\phi_0}{\partial x_j}
	+ f'(\phi_0)\frac{\partial^2\phi_0}{\partial x_k\partial x_j}\,,
}{auxId}
as follows
\bsplitl{
-{\cg B}_1
	&\ebrkts{
		f'(\phi_0)\phi_1
    	}
	=
	-\sum_{i,j=1}^d {\rm m}_{ij}\sum_{k=1}^d\frac{\partial\xi^k_\phi}{\partial y_j}
	\frac{\partial^2 f(\phi_0)}{\partial x_k\partial x_i}
\,.
}{B1}
After adding to \reff{B1} the term $-{\cal B}_2f(\phi_0)$, we can define a 
tensor $\hat{\rm M}_v=\brcs{{\rm m}^\phi_{ij}}_{1\leq i,k\leq d}$, i.e.,
\bsplitl{
{\rm m}^\phi_{ik}
    & :=
    \frac{1}{\av{Y}}\sum_{j=1}^d\int_{Y^1}\brkts{
	{\rm m}_{ik}
        -{\rm m}_{ij}\pd{\xi^k_\phi}{y_j}
    }\,d{\bf y}\,,
}{Mv}
such that
\bsplitl{
-{\cg B}_1
	\ebrkts{
		f'(\phi_0)\phi_1
    	}
	-{\cg B}_2f(\phi_0)
	= {\rm div}\brkts{
		\hat{\rm M}_\phi\nabla f(\phi_0)
	}
	\,.
}{B1B2}
%It remains to elucidate the last term in \reff{SoCow2}. Using \reff{O-0n}$_2$,
%then we have,
%\bsplitl{
%    -{\cg B}_0 \ebrkts{
%        \frac{1}{2}f''(\phi_0)\phi_1^2
%        +f'(\phi_0)\phi_2
%    }
%    =
%    \sum_{i,j=1}^d\pd{}{y_i}\brkts{
%        {\rm m}_{ij}\phi_1\pd{\phi_1}{y_j}
%    }f''(\phi_0)
%\\%&\quad
%    +\sum_{i,j=1}^d\pd{}{y_i}\brkts{
%        {\rm m}_{ij}\pd{\phi_2}{y_j}
%    }f'(\phi_0)
%\\
%    =
%    \sum_{i,j=1}^d-\ebrkts{
%        \pd{}{y_i}\brkts{
%            {\rm m}_{ij}\phi_1\pd{\xi^k_\phi}{y_j}
%        }
%    }f''(\phi_0)\pd{\phi_0}{x_k}
%    +{\rm m}\brkts{
%        {\cg A}_2\phi_0
%        +{\cg A}_1\phi_1
%        -w_0
%    }\,.
%}{B0Phi}
%where it is straightforward to conclude that \reff{B0Phi} is zero as an integral over $Y^1$.

These considerations finally lead to
the following effective equation for $\phi_0$, i.e.,
\bsplitl{
\theta_1\pd{\phi_0}{t}
    = {\rm div}\brkts{
	\hat{\rm M}_\phi/\lambda\nabla f(\phi_0)
    }
    +\frac{\lambda}{\theta_1}{\rm div}\brkts{
        \hat{\rm M}_w\nabla \brkts{
            {\rm div}\brkts{
                \hat{\rm D}\nabla \phi_0
            }
            -\tilde{g}_0
        }
    }\,.
}{EfW0}
The solvability of \reff{EfW0} follows along with the arguments in \cite{Novick-Cohen1990} 
because of Assumption A.
% at least assume
%that $f\in C^2_{Lip}(I)$ where $I\subset\mathbb{R}$ is a bounded interval. 
In fact, one immediately obtains a local Lipschitz continuity of the first two terms on the right hand side of \reff{EfW0}. For further details we refer the interested reader to Section \ref{sec:2Fo} \reff{sec:Ex} and
\cite{Novick-Cohen1990}.

\section{Proof of Theorem \ref{thm:ErEs}}\label{sec:ErEs}
For the derivation of the error estimates \reff{ThmErEs}, we work with the splitting formulation introduced in \cite{Schmuck2012b} and summarized in \reff{DePhMo0}. Hence, we compare the solution of the micrscopic porous media formulation \reff{PeMoPr} and with the solution of the effective homogenized porous media formulation \reff{pmWrThm}. 
As in \cite{Schmuck2012}, we introduce the error variables 
\bsplitl{
{\rm E}^w_\epsilon
	& := w^\epsilon-(w_0+(\epsilon w_1+\epsilon^2w_2))
	=: {\rm E}_0^w +\epsilon {\rm E}_1^w +\epsilon^2 {\rm E}_2^w
	\,,
\\
{\rm E}^\phi_\epsilon
	& := \phi^\epsilon-(\phi_0+(\epsilon\phi_1+\epsilon^2\phi_2))
	=: {\rm E}_0^\phi +\epsilon {\rm E}_1^\phi+\epsilon^2 {\rm E}_2^\phi
	\,,
\\ 
f^\epsilon(\phi^\epsilon,\phi_0,\phi_1,\phi_2)
	& := f(\phi^\epsilon)
	-\brkts{
		f(\phi_0)
		+\epsilon f'(\phi_0)\phi_1
		+\epsilon^2(f''(\phi_0)\phi_1^2+f'(\phi_0)\phi_2
	}
%	+{\cg O}(\epsilon^3)
%\\
%f^\epsilon({\rm E}^w_\epsilon)
%	& := f({\rm E}_0^w)
%		-\epsilon f'({\rm E}_0^w)w_1%+\epsilon w_2)
%	\,,
%%		+ \frac{\epsilon^2}{2}f''({\rm E}_0^w) (w_1+\epsilon w_2)^2
%%		+ {\cal O}(\epsilon^2)\,,
%\\
%f^\epsilon({\rm E}^\phi_\epsilon)
%	& := f({\rm E}_0^\phi)
%		-\epsilon f'({\rm E}_0^\phi)\phi_1%+\epsilon\phi_2)
%	\,,
%%		+ \frac{\epsilon^2}{2}f''({\rm E}_0^\phi)(\phi_1+\epsilon\phi_2)^2
%%		+ {\cal O}(\epsilon^2)\,,
	\,.
}{Vars}
The first goal is to determine the variable ${\rm F}_\epsilon^\iota$ which allows to 
write the equation for the errors ${\rm E}_\epsilon^\iota$ for $\iota\in\brcs{w,\phi}$ as follows
\bsplitl{
\begin{cases}
\quad
\frac{\partial {\cg A}_\epsilon^{-1}{\rm E}^w_\epsilon}{\partial t}
	= %{\rm div}\brkts{
	%	\hat{\rm M}\nabla 
		{\cg B}_\epsilon\ebrkts{ 
			-\lambda{\rm E}^w_\epsilon
			+ 
			\frac{1}{\lambda}f^\epsilon(\phi^\epsilon,\phi_0,\phi_1,\phi_2)
%			\brkts{
%				f(\phi^\epsilon)
%				-(f(\phi_0)+\epsilon f'(\phi_0)\phi_1
%			}
		}
	%}
	+ \epsilon {\rm F}^w_\epsilon
	& \textrm{in }\Omega^\epsilon\times ]0,T[
	\,,
\\\qquad\quad
\nabla_n{\rm E}^w_\epsilon
	= \epsilon {\rm G}^w_\epsilon
	& \textrm{on }\partial\Omega^\epsilon\times ]0,T[
	\,,
\\\quad
{\cg A}_\epsilon {\rm E}^\phi_\epsilon 
	= {\rm E}^w_\epsilon
	+ \epsilon {\rm F}^\phi_\epsilon
	& \textrm{in }\Omega^\epsilon\times ]0,T[
	\,,
\\\qquad\quad
\nabla_n{\rm E}^\phi_\epsilon
	= \epsilon {\rm G}^\phi_\epsilon
	& \textrm{on }\partial\Omega^\epsilon\times ]0,T[
	\,.
\end{cases}
}{ErEq}
We note that the second order terms $\phi_2$ and $w_2$ are used in the derivation 
of the error equation \reff{ErEq}. However, it is possible to derive error estimates which only 
depend on the first order correctors $\xi_\phi^k$ and $\xi_w^k$ for $1\leq k\leq d$.
With the definitions \reff{B0B1B2} we can rewrite the first term on the right-hand sides in \reff{ErEq}$_1$ and \reff{ErEq}$_2$ 
as follows
\bsplitl{
%{\rm div}\brkts{
%		\hat{\rm M}\nabla 
	{\cg B}_\epsilon
	\ebrkts{ 
			-\lambda{\rm E}^w_\epsilon
			+ 
			\frac{1}{\lambda}f^\epsilon(\phi^\epsilon,\phi_0,\phi_1,\phi_2)
			%f_\epsilon({\rm E}^\phi_\epsilon)		
		}
	%}
	& = -\brcs{
		\epsilon^{-2}{\cal B}_0
		+\epsilon^{-1}{\cal B}_1
		+{\cal B}_2
	}\ebrkts{
		-\lambda{\rm E}^w_\epsilon
			+ 
			\frac{1}{\lambda}f^\epsilon(\phi^\epsilon,\phi_0,\phi_1,\phi_2)
%			\brkts{
%				f(\phi^\epsilon)
%				-(
%					f(\phi_0)
%					+\epsilon f'(\phi_0)\phi_1
%				)
%			}
	}\,,
\\
{\cg A}_\epsilon {\rm E}_\epsilon^\phi
	& = -\brcs{
		\epsilon^{-2}{\cal A}_0
		+\epsilon^{-1}{\cal A}_1
		+{\cal A}_2
	}{\rm E}_\epsilon^\phi\,.
}{Re1TeRHS}
The relations in \reff{Re1TeRHS} together with the sequence of problems \reff{O-2n}, \reff{O-1n}, and \reff{O-0n} defines the terms ${\rm F}^w_\epsilon$ and ${\rm F}^\phi_\epsilon$ by
\bsplitl{
{\rm F}^w_\epsilon
	& := -\brcs{
			\lambda\brkts{
				{\cg B}_1w_2+ {\cal B}_2 w_1
			} 
			+\frac{1}{\lambda}\brkts{
				{\cg B}_1\ebrkts{
					f'(\phi_0)\phi_2
					+f''(\phi_0)\phi_1^2
				}
				+{\cal B}_2\ebrkts{f'({\rm E}_0^\phi)\phi_1}
			}
		}
\\&\quad\,\,
		-\epsilon\brcs{
			\lambda {\cg B}_2w_2
			+\frac{1}{\lambda}{\cg B}_2\ebrkts{
				f'(\phi_0)\phi_2
				+f''(\phi_0)\phi_1^2
			}
		}
	\,,
\\
{\rm F}^\phi_\epsilon
	& := -\brkts{
			{\cg A}_2 \phi_1
			+{\cg A}_1 \phi_2
		}
		-\epsilon%\brkts{
			{\cg A}_2\phi_2
		%}
		\,.
}{Fweps}
The inhomogeneities in the boundary conditions in \reff{ErEq} satisfy
\bsplitl{
{\rm G}^w_\epsilon
	& 
	:= -w_1-\epsilon w_2
	= \sum_{k=1}^d\xi_w^k\frac{\partial w_0}{\partial x_k}
	-\epsilon \sum_{k,l=1}^d\zeta_w^k\frac{\partial^2 w_0}{\partial x_k\partial x_l}
	\,,
\quad\textrm{and}\\
{\rm G}^\phi_\epsilon 
	&
	:= -\phi_1 -\epsilon \phi_2
	= \sum_{k=1}^d\xi_\phi^k\frac{\partial \phi_0}{\partial x_k}
	-\epsilon \sum_{k,l=1}^d\zeta_\phi^k\frac{\partial^2 \phi_0}{\partial x_k\partial x_l}
	\,,
}{GwGphi}
since the boundary conditions imply $\nabla_n(\iota^\epsilon-\iota_0)=0$ for 
$\iota\in\brcs{w,\phi}$. The second order correctors $\zeta^k_\phi$ and $\zeta^k_w$ are 
obtained from the reference cell problems \reff{O-0n} after using the ansatz 
\bsplitl{
\phi_2
	& = \sum_{k,l=1}^d\zeta_\phi^k\frac{\partial^2 \phi_0}{\partial x_k\partial x_l}
\quad\textrm{ and }\quad
w_2
	= \sum_{k,l=1}^d\zeta_w^k\frac{\partial^2 w_0}{\partial x_k\partial x_l}
	\,.
}{2ndCrr}
Elliptic theory allows us to estimate \reff{ErEq}$_3$ by
\bsplitl{
\N{{\rm E}^\phi_\epsilon}{H^1(\Omega)}
	& \leq 
		C\brkts{
			\N{{\rm E}^w_\epsilon}{L^2(\Omega)}
			+\epsilon \N{{\rm F}^\phi_\epsilon}{L^2(\Omega)}
			+\epsilon^{1/2}\N{G^\phi_\epsilon}{H^{-1/2}(\partial\Omega)}
		}
\\&
	\leq 
		C\N{{\rm E}^w_\epsilon}{L^2(\Omega)}
		+\epsilon C\brkts{1+\epsilon^{-1/2}}
	\,,
}{ClElEs}
where we subsequently justify the uniform boundedness of ${\rm F}^\phi_\epsilon$ 
in $L^2(\Omega)$.

It holds that
\bsplitl{
\N{{\rm F}^\phi_\epsilon}{L^2(\Omega)}
	\leq C\sum_{i,j,k,l=1}^d\N{\frac{\partial^3\phi_0}{\partial x_i\partial x_k\partial x_l}}{L^\infty}
		\brkts{
			\Ll{\delta_{ik}\xi^k_\phi\brkts{\frac{\cdot}{\epsilon}}}
			+\Ll{\delta_{ik}\zeta^k_\phi\brkts{\frac{\cdot}{\epsilon}}}
		}
	\leq C\,,
}{FphiL2}
for a constant $C>0$ independent of $\epsilon$. Analogously, we obtain the bound
\bsplitl{
\N{{\rm F}^w_\epsilon}{L^2(\Omega)}
&	\leq C\sum_{i,j,k,l=1}^d\N{\frac{\partial^3w_0}{\partial x_i\partial x_k\partial x_l}}{L^\infty}
		\biggl(
			\Ll{\delta_{ik}\xi^k_w\brkts{\frac{\cdot}{\epsilon}}}
			+\Ll{\delta_{ik}\zeta^k_w\brkts{\frac{\cdot}{\epsilon}}}
%\\&\quad
		+\frac{1}{\lambda}\Ll{
				{\cg B}_2\ebrkts{
					f'(\phi_0)\phi_2
					+f''(\phi_0)\phi_1^2
				}
		}
\\&\quad
		+\frac{1}{\lambda}\brkts{
				\Ll{{\cg B}_1\ebrkts{
					f'(\phi_0)\phi_2
					+f''(\phi_0)\phi_1^2
				}}
				+\Ll{{\cal B}_2\ebrkts{f'({\rm E}_0^\phi)\phi_1}}
		}
		\biggr)
		\leq C\,,
}{FwL2}
where we used the fact that $f(s)$ is a polynomial of order $2p-1$ by Assumption (PC), i.e.,
\bsplitl{
\av{f'(s)}
	&\leq c\brkts{
		1+\av{s}^{2p-2}
	}
	\,,
\qquad%\textrm{ and }\quad
\av{f''(s)}
	\leq c\brkts{
		1+\av{s}^{2p-3}
	}
	\,,
\quad\textrm{ and }\quad
\av{f'''(s)}
	\leq c\brkts{
		1+\av{s}^{2p-4}
	}
	\,.
}{Polyf}
In order to control ${\rm G}_\epsilon^\phi$, we apply a standard argument \cite{Bensoussans1978,Cioranescu1999} based on a cut-off function $\chi^\epsilon$ which is defined as follows,
\bsplitl{
\begin{cases}
\quad
\chi^\epsilon \in {\cg D}(\Omega)\,,&
\\\quad
\chi^\epsilon = 1
	& \textrm{if }{\rm dist}(x,\partial\Omega)\leq\epsilon\,,
\\\quad
\chi^\epsilon = 0
	& \textrm{if }{\rm dist}(x,\partial\Omega)\geq 2 \epsilon\,,
\\\quad
\N{\nabla \chi^\epsilon}{L^\infty(\Omega)}
	\leq \frac{C}{\epsilon}\,.
\end{cases}
}{CF}
For $\eta^\phi_\epsilon :=  \chi^\epsilon {\rm G}^\phi_\epsilon$ we show that $\eta^\phi_\epsilon\in H^1(\Omega)$ and 
\bsplitl{
\N{\eta^\phi_\epsilon}{H^1(U^\epsilon)}
	\leq C\epsilon^{-1/2}\,,
}{G3Best}
where $U^\epsilon$ is the support of $\eta^\phi_\epsilon$ and forms a neighbourhood of $\partial\Omega$ of thickness $2\epsilon$. By the regularity properties of $\xi^k_w$, 
$\xi_\phi^k$ and $\chi^\epsilon$ immediately obtain the bound
\bsplitl{
\N{\eta^\phi_\epsilon}{H^1(U^\epsilon)}
	\leq C\brkts{
		\frac{1}{\epsilon}\N{\phi_0}{H^1(U^\epsilon)}
		+1
	}
	\,,
}{grdEta}
for a $C$ independent of $\epsilon$. Next we use the result (\cite[Lemma 5.1, p.7]{Oleinik1992}), that is,
%
%\bsplitl{
$
\N{\phi_0}{H^1(U^\epsilon)}
	\leq \epsilon^{1/2}C\N{\nabla\phi_0}{H^1(\Omega)}
	\,,
$
%}{Ol92}
%
for a $C$ independent of $\epsilon$. Herewith, we established \reff{G3Best}. Using the 
trace theorem and the fact that $\eta^\phi_\epsilon=G^\epsilon$ on $\partial\Omega$ allow 
us to obtain the following estimate
\bsplitl{
\N{{\rm G}^\phi_\epsilon}{H^{1/2}(\partial\Omega)}
	= \N{\eta^\phi_\epsilon}{H^{1/2}(\partial\Omega)}
	\leq C\N{\eta^\phi_\epsilon}{H^1(\Omega)}
	= C\N{\eta^\phi_\epsilon}{H^1(U^\epsilon)}
	\,,
}{GepsCtrl}
which provides with \reff{G3Best} the bound
\bsplitl{
\N{{\rm G}_\epsilon^\phi}{H^{1/2}(\partial\Omega)}
	\leq C\epsilon^{-1/2}
	\,.
}{GepsCtrl1}
Applying the same arguments to ${\rm G}_\epsilon^w$ immediately leads to 
the corresponding bound
\bsplitl{
\N{{\rm G}_\epsilon^w}{H^{1/2}(\partial\Omega)}
	\leq C\epsilon^{-1/2}
	\,.
}{GWepsCtrl}

Next, we estimate \reff{ErEq}$_1$.
%%%%%%%%%%%%% Definitions for a cut-off function %%%%%%%%%%%%%%%%%
%[
%where $\chi_\epsilon\in C^\infty_0(\Omega)$ is a cut-off function satisfying 
%$0\leq\chi_\epsilon\leq 1$, $\epsilon\av{\nabla\chi_\epsilon}\leq C$, and 
%$\chi_\epsilon=1$ on 
%$\brcs{{\bf x}\in\Omega\,\bigl|\,{\rm dist}({\bf x},\partial\Omega)\geq\epsilon}$. 
%It further holds that $\Ll{\nabla\chi_\epsilon}\leq C\epsilon^{-1/2}$ 
%and $\Ll{1-\chi_\epsilon}\leq C\epsilon^{1/2}$ \cite{Bensoussans1978,Cioranescu1999}.
%We note that the use of the cut-off function $\chi_\epsilon$ simplifies the subsequent 
%analysis such that we do not have to account for inhomogeneous boundary 
%terms/conditions.
%]
%%%%%%%%%%%%%%%%%%%%%%%%%%%%%%%%%%%%%%%%%%%%%%%%
Testing \reff{ErEq}$_1$ with ${\cg A}_\epsilon {\rm E}_\epsilon^w$ provides the 
equations
\bsplitl{
\brkts{
		\partial_t {\cg A}_\epsilon^{-1}{\rm E}_\epsilon^w, {\cg A}_\epsilon {\rm E}_\epsilon^w
	}
	& = \brkts{
		\partial_t {\rm E}^w_\epsilon,{\rm E}_\epsilon^w
	}
	+\int_{\partial\Omega}\brcs{
		{\rm E}_\epsilon^\phi-\epsilon{\cg A}_\epsilon^{-1}{\rm F}_\epsilon^\phi
	}{\rm G}_\epsilon^w\,d{\bf x}
	\,,
\\
-\lambda\brkts{
		{\cg B}_\epsilon {\rm E}_\epsilon^w,{\cg A}_\epsilon {\rm E}_\epsilon^w
	}
	& = -\lambda m\brkts{
		{\cg A}_\epsilon{\rm E}_\epsilon^w,
		{\cg A}_\epsilon{\rm E}_\epsilon^w
	}
	= -\lambda m\Ll{{\cg A}_\epsilon{\rm E}_\epsilon^w}^2
	\,,
\\
\brkts{
		{\cg B}_\epsilon f^\epsilon(\phi^\epsilon,\phi_0,\phi_1,\phi_2),
		{\cg A}_\epsilon{\rm E}^w_\epsilon
	}
	& =
	m \Bigl(
		{\cg A}_\epsilon\Bigl\{
			f(\phi^\epsilon)-f(\phi_0)
			-\epsilon f'(\phi_0)\phi_1
\\&\quad
			-\epsilon^2(
				f'(\phi_0)\phi_2+f''(\phi_0)\phi_2^2
			)
		\Bigr\},
		{\cg A}_\epsilon{\rm E}_\epsilon^w
	\Bigr)
	\,,
}{Eweps1}
which together lead to the estimate
\bsplitl{
\frac{1}{2}\frac{d}{dt}\Ll{{\rm E}^w_\epsilon}^2
	+m\lambda\Ll{{\cg A}_\epsilon{\rm E}_\epsilon^w}^2
	& \leq 
	C(m,\lambda)\Bigl(
		\av{\brkts{
			{\cg A}_\epsilon\ebrkts{
				f(\phi^\epsilon)
				-f(\phi_0)
			},{\cg A}_\epsilon{\rm E}_\epsilon^w
		}}
%\pmb{		\av{\brkts{
%			{\cg A}_\epsilon f(\phi^\epsilon),
%			{\cg A}_\epsilon {\rm E}_\epsilon^w
%		}}
%		+\av{\brkts{
%			{\cg A}_\epsilon f(\phi_0),
%			{\cg A}_\epsilon {\rm E}_\epsilon^w
%		}}
%}
%
		+\epsilon\av{\brkts{
			{\cg A}_\epsilon[f'(\phi_0)\phi_1],
			{\cg A}_\epsilon{\rm E}_\epsilon^2
		}}
		+{\cg O}(\epsilon^2)
	\Bigr)
\\&\quad
	+\epsilon\int_{\partial\Omega}\av{
		\brcs{
			{\rm E}_\epsilon^\phi
			-\epsilon{\cg A}_\epsilon^{-1}{\rm F}_\epsilon^\phi
		}{\rm G}_\epsilon^w
	}\,d{\bf x}
	\,.
}{Eweps2}
In order to control the terms on the right-hand side in \reff{Eweps2}, we make use of 
the fact that $f(s)$ is a polynomial and satisfies \reff{Polyf}.
%Herewith, the first term on the right-hand side of \reff{Eweps2} can be estimated by
%%
%\bsplitl{
%\av{\brkts{
%			{\cg A}_\epsilon f(\phi^\epsilon),
%			{\cg A}_\epsilon {\rm E}_\epsilon^w
%	}}
%&	\leq c\av{\Bigl(\brkts{
%		1+\av{\phi^\epsilon}^{2p-3}
%	}\av{\nabla\phi^\epsilon}^2,
%	{\cg A}_\epsilon{\rm E}_\epsilon^w
%	\Bigr)}
%	+c\av{\Bigl(\brkts{
%		1+\av{\phi^\epsilon}^{2p-2}
%	}\av{\Delta\phi^\epsilon},
%	{\cg A}_\epsilon{\rm E}_\epsilon^w
%	\Bigr)}
%\\&
%	\leq 
%	c\brkts{
%		1+\N{\phi^\epsilon}{L^\infty}^{2p-3}
%	}
%	\N{\nabla\phi^\epsilon}{L^4}^2
%	\Ll{{\cg A}_\epsilon{\rm E}_\epsilon^w}
%	+c\brkts{
%		1+\N{\phi^\epsilon}{L^\infty}^{2p-2}
%	}
%	\Ll{\Delta\phi^\epsilon}
%	\Ll{{\cg A}_\epsilon{\rm E}_\epsilon^w}
%\\&
%	\leq c(\kappa)\brkts{
%		\brkts{
%			1+\N{\phi^\epsilon}{L^\infty}^{2p-3}
%		}^2
%		\N{\nabla\phi^\epsilon}{L^4}^4
%		+\brkts{
%			1+\N{\phi^\epsilon}{L^\infty}^{2p-2}
%		}^2
%		\Ll{\Delta\phi^\epsilon}^2
%	}
%	+\kappa\Ll{{\cg A}_\epsilon{\rm E}_\epsilon^w}^2
%	\,.
%}{1Eweps2}
%%
%It is immediately apparent that the analogous estimate holds for the second term in 
%\reff{Eweps2}. 
%\\
%{\bf The above estimates should be replaced by:}
%\\
The first term in \reff{Eweps2} satisfies the following inequality
\bsplitl{
%\pmb{
\av{\brkts{
		{\cg A}_\epsilon\ebrkts{
			f(\phi^\epsilon)
			-f(\phi_0)
		},{\cg A}_\epsilon{\rm E}_\epsilon^w
	}}
%}
&	\leq C\Bigl(
	\av{\brkts{
		\brcs{
			f''(\phi^\epsilon)
			-f''(\phi_0)
		}\av{\nabla\phi^\epsilon}^2,
		{\cg A}_\epsilon{\rm E}_\epsilon^w
	}}
\\&\quad
	+\av{\brkts{
		f''(\phi_0)\nabla\brkts{
			\phi^\epsilon
			+\phi_0
		}\nabla\brkts{
			\phi^\epsilon
			-\phi_0
		},
		{\cg A}_\epsilon{\rm E}_\epsilon^w
	}}
\\&\quad
	+\av{\brkts{
		\brcs{
			f'(\phi^\epsilon)
			-f'(\phi_0)
		}\Delta\phi^\epsilon,
		{\cg A}_\epsilon{\rm E}_\epsilon^w
	}}
\\&\quad
	+\av{\brkts{
		f'(\phi_0)\Delta(\phi^\epsilon-\phi_0),
		{\cg A}_\epsilon{\rm E}_\epsilon^w
	}}
	\Bigr)
\,.
}{FrEnDi}
Before we proceed, we estimate the terms on the right-hand side in \reff{FrEnDi}:\\
\emph{1st term in \reff{FrEnDi}:} We first note that with the remainder term in Taylor series 
we obtain 
\bsplitl{
\av{f''(\phi^\epsilon)
		-f''(\phi_0)
	}
&	\leq \av{
		\sup_{\theta\in I_\phi}f'''(\theta)\brkts{
			\phi^\epsilon-\phi_0
		}
	}
	\leq T\av{\Omega}
		\N{f'''(\theta)}{L^\infty(I_\phi)}
		\av{\phi^\epsilon-\phi_0}
\\&
	\leq T\av{\Omega}
		\N{f'''(\theta)}{L^\infty(I_\phi)}
		\brkts{
			\av{{\rm E}_\epsilon^\phi}
			+\epsilon\N{\phi_1}{L^\infty(\Omega_T)}
			+\epsilon^2\N{\phi_2}{L^\infty(\Omega_T)}
		}
	\,,
}{1FrEnDi}
where $I_\phi:=]\underline{\phi},\overline{\phi}[$ is the interval defined by the smallest 
real root $\underline{\phi}$ and $\overline{\phi}$ the largest real root of the polynomial free energy 
$F$ characterized by \reff{PolyDef}.
Herewith, we can estimate the first term (e.g. in $d=3$) as follows
\bsplitl{
&
\av{\brkts{
		\brcs{
			f''(\phi^\epsilon)
			-f''(\phi_0)
		}\av{\nabla\phi^\epsilon}^2,
		{\cg A}_\epsilon{\rm E}_\epsilon^w
	}}
	\leq C(\Omega,T)\N{f'''(\theta)}{L^\infty(I_\phi)}
		\N{{\rm E}_\epsilon^\phi}{L^4}
		\N{\nabla\phi^\epsilon}{L^6}
		\Ll{{\cg A}_\epsilon {\rm E}_\epsilon^w}
\\&\qquad\qquad\qquad\qquad\qquad\qquad
	\leq C(\Omega,T)\N{f'''(\theta)}{L^\infty(I_\phi)}
		\N{{\rm E}_\epsilon^\phi}{H^1}
		\N{\nabla\phi^\epsilon}{H^1}
		\Ll{{\cg A}_\epsilon {\rm E}_\epsilon^w}
\\&\qquad\qquad\qquad\qquad\qquad\qquad
	\leq  C(\Omega,T,\kappa)\N{{\rm E}_\epsilon^\phi}{H^1}^2
		+\kappa\Ll{{\cg A}_\epsilon {\rm E}_\epsilon^w}^2
\\&\qquad\qquad\qquad\qquad\qquad\qquad 
	\leq C(\Omega,T,\kappa) \brkts{
			\Ll{{\rm E}_\epsilon^w}^2
			+ \epsilon^2 (1+\epsilon^{1/2})^2
		}
		+\kappa\Ll{{\cg A}_\epsilon {\rm E}_\epsilon^w}^2
	\,.
}{1FrEnDi1}
\\
\emph{2nd term in \reff{FrEnDi}:} With Sobolev inequalities and the identity 
$\phi^\epsilon-\phi_0={\rm E}_\epsilon^\phi+\epsilon\phi_1+\epsilon^2\phi_2$ we 
obtain the following estimate
\bsplitl{
\av{\brkts{
		f''(\phi_0)\nabla\brkts{
			\phi^\epsilon
			+\phi_0
		}\nabla\brkts{
			\phi^\epsilon
			-\phi_0
		},
		{\cg A}_\epsilon{\rm E}_\epsilon^w
	}}
&	\leq
		\N{f'''(\cdot)}{L^\infty(I_\phi)}
		\brkts{
			\N{\nabla\phi^\epsilon}{L^6}
			+\N{\nabla\phi_0}{L^6}
		}\Bigl(
			\N{\nabla{\rm E}_\epsilon^\phi}{L^3}
\\&\quad
			+\epsilon\N{\nabla\phi_1}{L^3}
			+\epsilon^2\N{\nabla\phi_2}{L^3}
		\Bigr)
		\Ll{{\cg A}_\epsilon {\rm E}_\epsilon^w}
\\&
	\leq C(\Omega,T,\kappa) \Bigl(
			\Ll{{\rm E}_\epsilon^w}^2
			+ \epsilon^2\brkts{ 
				(1+\epsilon^{1/2})^2
				+(1+\epsilon^2)
			}
		\Bigr)
		+\kappa\Ll{{\cg A}_\epsilon {\rm E}_\epsilon^w}^2
	\,.
}{2FrEnDi}
\\
\emph{3rd term in \reff{FrEnDi}:} Following the same ideas as for the \emph{1st term} estimated 
in \reff{1FrEnDi1}, we immediately get the bound
\bsplitl{
\av{\brkts{
		\brcs{
			f'(\phi^\epsilon)
			-f'(\phi_0)
		}\Delta\phi^\epsilon,
		{\cg A}_\epsilon{\rm E}_\epsilon^w
	}}
&	\leq C(\Omega,T,\kappa) \brkts{
			\Ll{{\rm E}_\epsilon^w}^2
			+ \epsilon^2 (1+\epsilon^{1/2})^2
		}
		+\kappa\Ll{{\cg A}_\epsilon {\rm E}_\epsilon^w}^2
	\,.
}{3FrEnDi}
\\
\emph{4th term in \reff{FrEnDi}:} The last term can finally be controlled as follows
\bsplitl{
\av{\brkts{
		f'(\phi_0)\Delta(\phi^\epsilon-\phi_0),
		{\cg A}_\epsilon{\rm E}_\epsilon^w
	}}
&	\leq
	\N{f''(\cdot)}{L^\infty}\Bigl(
		\Ll{\Delta{\rm E}_\epsilon^\phi}
		+\epsilon\Ll{\Delta\phi_1}
		+\epsilon^2\Ll{\Delta\phi_2}
	\Bigr)\Ll{{\cg A}_\epsilon{\rm E}_\epsilon^w}
\\&
	\leq 
	C(\Omega,T,\kappa)\N{f''(\cdot)}{L^\infty}\Bigl(
		\brkts{
			\Ll{{\rm E}_\epsilon^w}^2
			+\epsilon^2\Ll{{\rm F}_\epsilon^\phi}^2
		}
		+\epsilon^2%\Ll{\Delta\phi_1}^2
		+\epsilon^4%\Ll{\Delta\phi_2}^2
	\Bigr)+\kappa\Ll{{\cg A}_\epsilon{\rm E}_\epsilon^w}^2
	\,.
}{4FrEnDi}
\\
The third term in \reff{Eweps2} can be controlled by
\bsplitl{
\epsilon\av{\brkts{
		{\cg A}_\epsilon[f'(\phi_0)\phi_1],
		{\cg A}_\epsilon{\rm E}_\epsilon^w
	}}
&	\leq \epsilon c \Bigl(
		\brkts{1+\N{\phi_0}{L^\infty}^{2p-4}}\N{\nabla\phi_0}{L^4}^2\av{\phi_1}
		+\brkts{
			1+\N{\phi_0}{L^\infty}^{2p-3}
		}\Bigl\{
			\N{\nabla\phi_1}{L^3}\N{\nabla\phi_0}{L^6}
\\&\quad
			+\N{\phi_1}{L^\infty}\Ll{\Delta\phi_0}
		\Bigr\}
	\Bigr)
	\Ll{{\cg A}_\epsilon{\rm E}_\epsilon^w}
\\&\leq
	\epsilon^2 c(\kappa)\Bigl(
		\brkts{1+\N{\phi_0}{L^\infty}^{2p-4}}^2\N{\nabla\phi_0}{L^4}^4\N{\phi_1}{L^\infty}^2
\\&\quad
		+\brkts{
			1+\N{\phi_0}{L^\infty}^{2p-3}
		}^2\Bigl\{
			\N{\nabla\phi_1}{L^3}^2\N{\nabla\phi_0}{L^6}^2
%\\&\quad
			+\N{\phi_1}{L^\infty}^2\Ll{\Delta\phi_0}^2
		\Bigr\}
	\Bigr)
	+\kappa
	\Ll{{\cg A}_\epsilon{\rm E}_\epsilon^w}^2
	\,.
}{3Eweps2}
It leaves to bound the last term in \reff{Eweps2}. Again using the cut-off 
function $\chi^\epsilon$ introduced in \reff{CF} and definining 
$\rho_\epsilon^\phi:=\chi^\epsilon{\rm G}_\epsilon^\phi$ enables us to 
derive the analogous inequality 
$\N{\rho_\epsilon^\phi}{H^1(U^\epsilon)}\leq C\epsilon^{-\frac{1}{2}}$ as in 
\reff{G3Best} such that with the trace theorem the following bound holds 
\bsplitl{
\N{{\rm E}_\epsilon^\phi}{H^{1/2}(\partial\Omega)}
	=\N{\rho_\epsilon^\phi}{H^{1/2}(\partial\Omega}
	\leq C\N{\rho_\epsilon^\phi}{H^1(\Omega)}
	=C\N{\rho_\epsilon^\phi}{H^1(U^\epsilon)}
	\leq C\epsilon^{-\frac{1}{2}}
	\,.
}{EphiEps}
Herewith, we can estimate the last term in \reff{Eweps2} as follows
\bsplitl{
\epsilon\int_{\partial\Omega}\av{
		\brcs{
			{\rm E}_\epsilon^\phi
			-\epsilon{\cg A}_\epsilon^{-1}{\rm F}_\epsilon^\phi
		}{\rm G}_\epsilon^w
	}\,do
&	\leq \epsilon\brkts{
		\N{{\rm E}^\phi_\epsilon}{H^{-\frac{1}{2}}(\partial\Omega)}
		+\epsilon \N{{\cg A}_\epsilon^{-1}{\rm F}^\phi_\epsilon}{H^{-\frac{1}{2}}(\partial\Omega)}
		}
		\N{{\rm G}^w_\epsilon}{H^{1/2}(\partial\Omega)}
\\&
	\leq \epsilon C\brkts{
		\N{{\rm E}_\epsilon^\phi}{H^{1/2}(\partial\Omega)}
		+\epsilon
	}\epsilon^{1/2}
\\&
	\leq C\brkts{
		\epsilon
		+\epsilon^{5/2}
	}
	\,.
}{Bndry}
Putting things together finally allows us to rewrite the estimate \reff{Eweps2} 
as follows
\bsplitl{
\Ll{{\rm E}_\epsilon^w(\cdot,T)}^2
	+2\brkts{
		m\lambda-5\kappa
	}
	\int_0^T\Ll{{\cg A}_\epsilon{\rm E}_\epsilon^2}^2\,dt
	\leq C\brkts{\epsilon+\epsilon^{5/2}}
		\int_0^T{\rm exp}\brkts{B(T)-B(t)}\,dt
	\,,
}{Eweps3}
where $B(t):=\int_0^tC(\Omega,s,\kappa)\,ds$.

\section{Conclusions}\label{sec:Concl}
Based on a microscopic porous media formulation \reff{PeMoPr}, we formally
derived upscaled/homogenized phase field equations for arbitrary free energy
densities. We gave a rigorous justification of this new effective macroscopic
formulation for general polynomial free energies which include the widely used
double-well potential. The porous materials considered here can be
represented by a periodic covering of a single reference cell $Y$, which
takes reliably the pore geometry into account. It is well-known that
transport as well as fluid flow in porous media lead to high-dimensional
computational problems, since the mesh size needs to be much smaller than the
heterogeneity $\epsilon:=\frac{\ell}{L}$ which is the quotient of the
characteristic length scale of the pores $\ell$ and the size of macroscopic
porous medium $L$.

We note that many synthetically produced (e.g. by a silicon template
technique \cite[Section 1.2, p. 7]{Barsukov2006}) or commercially available
porous media show periodic heterogeneities such that the periodicity
assumption is for many applications realistic. We rigorously derived
qualitative error estimates for the approximation error between the solution
of the effective macroscopic problem \reff{pmWrThm} and the solution of the
fully resolved microscopic equation \reff{PeMoPr}. We also recovered the
widely known error behavior from homogenization of elliptic
problems %\cite{Cioranescu1999}
in the context of phase field problems for instance.

This error qunatification is of fundamental interest in applications as it gives guidance on the
applicability of the new effective macroscopic phase field formulation in dependence of the
heterogeneity $\epsilon$ of the considered porous medium. This dimensionally reduced
phase field formulation provides a reliable and convenient computational strategy where the
details of a single characteristic pores enter in a reliably averaged way preventing a 
full numerical resolution. Hence, one is only left with discretizing
the macroscopic porous medium.

Due to the popularity and the wide range of applicability of phase field equations, the
new, here rigorously derived, effective macroscopic formulation, serves as a promising
and valuable tool in material, chemical, physical sciences and engineering. However, there are many interesting open questions of major physical
interest. For instance: Are there restricting conditions under which this new formulation also 
reliably describes phase transformations in heterogeneous media such as composites and 
porous materials?
%Would it be possible to identify the
%upscaled/homogenized Hele-Shaw problem for perforated domains after establishing such a sharp
%interface limit for the here derived upscaled equation \reff{pmWrThm}?
Moreover, the error estimates do not seem to be optimal with respect to time.
The natural question then is whether the asymptotic behavior in the limit
$\epsilon\to 0$ is uniform in time.
%We note that the homogenization
%of the Hele-Shaw problem in a slightly different context has been studied in \cite{Kim2009}. However,
%the microscopic Hele-Shaw problem from \cite{Kim2009} would be different for our porous media case and
%as a result one the upscaled/homogenized equations will be different.
Finally, we believe that the effective phase field formulation \reff{pmWrThm}
serves as a promising new direction for modelling multiphase flow in porous
media by not only extending Darcy's law. We shall examine there and related
questions in future studies.

\section*{Acknowledgements}
We acknowledge financial support from EPSRC Grant No. EP/H034587, EPSRC Grant No. EP/J009636/1, and ERC Advanced Grant No. 247031.

          % BibTeX users please use
\bibliographystyle{plain} %{rspublicnat}  % rspublicnat} %plain
\bibliography{effWetting7} %stoMoReByRG_new,
          %
          % Non-BibTeX users please use

\end{document}